\documentclass[a4paper,twocolumn,11pt,accepted=2024-03-10]{quantumarticle}

\pdfoutput=1

\usepackage{amsmath,amssymb}
\usepackage{graphicx}
\usepackage{dcolumn}
\usepackage{bm}
\usepackage{xcolor}
\usepackage{multirow}
\usepackage{tabularx}
\usepackage{array}
\usepackage{booktabs}
\usepackage{upgreek}
\usepackage{makecell} 
\usepackage{ulem}
\usepackage[numbers,sort&compress]{natbib}

\newcommand{\ket}[1]{\left\vert{#1}\right\rangle}

\newcommand{\braket}[1]{\left\langle{#1}\right\rangle}
\newcommand{\parens}[1]{\left({#1}\right)}
\newcommand{\eqr}[1]{Eq.\ (\ref{#1})}

\newcommand{\de}{\delta \varepsilon}
\newcommand{\kt}{k_B T}

\usepackage{textcomp}
\usepackage{xcite}
\usepackage{verbatim}
\usepackage[colorinlistoftodos]{todonotes}
\usepackage{microtype}

\usepackage{xr-hyper}
\usepackage{hyperref}
\makeatletter
\newcommand*{\addFileDependency}[1]{
  \typeout{(#1)}
  \@addtofilelist{#1}
  \IfFileExists{#1}{}{\typeout{No file #1.}}
}
\makeatother

\newcommand{\appropto}{\mathrel{\vcenter{
  \offinterlineskip\halign{\hfil$##$\cr
    \propto\cr\noalign{\kern2pt}\sim\cr\noalign{\kern-2pt}}}}}

\begin{document}


\title{Beyond-adiabatic Quantum Admittance of a Semiconductor Quantum Dot at High Frequencies: Rethinking Reflectometry as Polaron Dynamics}  

\author{L. Peri}
\email{lp586@cam.ac.uk}
\affiliation{Cavendish Laboratory, University of Cambridge, J.J. Thomson Avenue, Cambridge CB3 0HE, United Kingdom}
\affiliation{Quantum Motion, 9 Sterling Way, London N7 9HJ, United Kingdom}
\author{G. A. Oakes}
\affiliation{Cavendish Laboratory, University of Cambridge, J.J. Thomson Avenue, Cambridge CB3 0HE, United Kingdom}
\affiliation{Quantum Motion, 9 Sterling Way, London N7 9HJ, United Kingdom}
\author{L. Cochrane}
\affiliation{Cavendish Laboratory, University of Cambridge, J.J. Thomson Avenue, Cambridge CB3 0HE, United Kingdom}
\affiliation{Quantum Motion, 9 Sterling Way, London N7 9HJ, United Kingdom} 
\author{C. J. B. Ford}
 \affiliation{Cavendish Laboratory, University of Cambridge, J.J. Thomson Avenue, Cambridge CB3 0HE, United Kingdom}
\author{M. F. Gonzalez-Zalba}
 \email{fernando@quantummotion.tech}
 \affiliation{Quantum Motion, 9 Sterling Way, London N7 9HJ, United Kingdom}

\date{August 2023}

\begin{abstract}
 Semiconductor quantum dots operated dynamically are the basis of many quantum technologies such as quantum sensors and computers. Hence, modelling their electrical properties at microwave frequencies becomes essential to simulate their performance in larger electronic circuits. Here, we develop a self-consistent quantum master equation formalism to obtain the admittance of a quantum dot tunnel-coupled to a charge reservoir under the effect of a coherent photon oscillator. We find a general expression for the admittance that captures the well-known semiclassical (thermal) limit, along with the transition to lifetime and power broadening regimes due to the increased coupling to the reservoir and amplitude of the photonic drive, respectively. Furthermore, we describe two new photon-mediated regimes: Floquet broadening, determined by the dressing of the QD states, and broadening determined by photon loss in the system. Our results provide a method to simulate the high-frequency behaviour of QDs in a wide range of limits, describe past experiments, and propose novel explorations of QD-photon interactions.      
\end{abstract}

\maketitle

\section{Introduction}
Semiconductor quantum dots (QDs) are a promising platform for developing solid-state quantum technologies. In the area of quantum computing~\cite{loss1998quantum}, six-qubit quantum processors~\cite{philips2022universal} and the fabrication of two-dimensional arrays of 4$\times$4 quantum dots have been shown~\cite{borsoi2022shared}. In conjunction, demonstrations of qubit control at the threshold for fault-tolerant computing~\cite{xue2022quantum,noiri2022fast,mills2022two}, advanced manufacturing~\cite{Maurand2016,zwerver2022qubits}, and integration with classical semiconductor electronics~\cite{xue2021cmos,ruffino2022cryo} indicate that QDs are a compelling platform for quantum computation.

More recently, new electronic applications of QD devices are beginning to emerge, particularly when manipulated at high frequencies. Their non-linear admittance, consisting of a combination of circuit equivalents such as the Sisyphus resistance~\cite{Petersson2010} and quantum and tunnelling capacitances~\cite{Vigneau_2023}, can be utilised for quantum sensing --- in the form of fast electrometers~\cite{house2016, oakes_fast_PRX,vanHeijden_Simmons_Rogge_2018} and local thermometers~\cite{Ahmed2018,chawner2021nongalvanic} --- and for electronic signal conversion --- in the form of frequency multipliers~\cite{oakes2022quantum} and quantum amplifiers~\cite{Cochrane2022}.

As these technologies become established, developing accurate electrical models of semiconductor QDs at high frequencies becomes essential to simulate their performance as stand-alone elements in hybrid quantum-classical circuits or circuit quantum-electrodynamics architectures~\cite{Mi2017, Samkharadzeeaar2018,Hansen_2022,Seedhouse_2021}. Semiclassical models accurately predict the behaviour in the limit of slow and weak driving, along with weak coupling to the environment~\cite{Mizuta2017, Esterli2019}. Furthermore, quantum models have been developed that further describe the admittance in the limit of strong coupling to the electron bath~\cite{Cottet2011,French2022}. However, a consolidated model that accurately describes the interaction of a QD with a charge reservoir and the quantum nature of the photons across multiple regimes is missing. 

Here, we present a Markovian Master Equation formalism for the charge dynamics in a driven QD tunnel-coupled to a charge reservoir (CR), hereafter, a semiconductor-based Single Electron Box (SEB) (see Fig.~\ref{fig:fig1}a) to obtain a general form of the effective quantum admittance of the SEB. The entire system can be described from a quantum mechanical point of view as three interacting subsystems: a QD with a single discrete electrochemical level, a CR in thermal equilibrium at temperature $T$, and a single-frequency Photonic bath, modelled as an Oscillator (PhO) representing microwave radiation. The QD can exchange charged particles with the CR, and both the photon number and frequency can be controlled by external means. 
We describe the role of temperature and charge tunnel rate to the reservoir with a smooth transition from thermal to lifetime broadening. Furthermore, we capture the impact of increasing the driving amplitude, resulting in power broadening of the charge transition. Our description of the coupled electron-photon states in the QD as a polaron will shed new and physically insightful light on this process.
Moreover, we present a novel kind of broadening, Floquet Broadening, which is dictated by the photonic part of the polaron. The Heisenberg uncertainty principle predicts this effect, but its intrinsically quantum nature makes it inaccessible to a simple extension of semiclassical adiabatic theories.
Finally, we describe the effects due to non-idealities in the PhO and how this affects the equivalent admittance of the SEB. This result is the direct analogue of known phenomena in superconductor transmon qubits \cite{gambetta_quantum_2008,gambetta_qubit-photon_2006,slichter_measurement_induced_2012}, and it shows how the formalism developed here lays the foundations to bridge the gap between equivalent circuit simulation and circuit quantum electrodynamics with mixed bosonic and fermionic states.
In particular, we identify five quantities that mostly dictate the SEB dynamics, which give rise to the same number of possible regimes. Because of the nature of the SEB admittance at the fundamental frequency, which takes the form of a single peak, we distinguish these regimes by the limiting factor that causes a broadening of the lineshape. (1) Thermal Broadening (TB) occurs when the peak is limited by the charge temperature of the CR. (2) Lifetime Broadening (LB) is where the broadening is due to the finite lifetime of the charge in the QD. (3) Floquet Broadening (FB), a high-frequency broadening caused by the discrete nature of the photon energy. (4) Power Broadening (PB) is a large-signal effect where the peak is dictated by a large amplitude of the drive. (5) Photon Loss Broadening (PLB), where the large-signal response is modified by a finite rate of photon loss in the (non-ideal) photonic oscillator.

\section{Statement of Key Results}

In this section, we introduce the necessary concepts and notation and summarize the key finding of this work while providing the required context and presenting physical implications.
Section~\ref{SEB_dynamics} describes the SEB and its time evolution by framing the charge dynamics quantum mechanically and proves that it only depends on the QD-CR tunnel rates.
Section~\ref{Tunnel_Rates} outlines the tunnel rates semiclassically and in a fully quantum setting. In the process, we will develop a formalism for a self-consistent quantum master equation of the SEB, which is general to other quantum systems embedded in classical circuits, for which it is interesting to consider effective circuit analogues.
These results are used to derive the equivalent SEB admittance, which is discussed from an electrical and physical point of view. While the formalism here developed is capable of capturing strong QD-PhO coupling effects, the discussion is carried forward only in the weak coupling regime. The strong QD-CR coupling case is however fully discussed.
Finally, Section~\ref{Regimes} presents the equivalent SEB admittance and its reflectometry signals, exploring the roles of the various physical and dynamic properties of the system. 
The discussion identifies several regimes of SEB operation, depending on the parameter dominating the charge dynamics, casting new insight into previous well-known results and identifying new effects when the semiclassical approximations are no longer valid.

In Sections~\ref{SEB_dynamics} and \ref{Tunnel_Rates}, we focus on the physical intuition behind the results presented in this work. The reader interested in a rigorous derivation and the mathematical methods employed therein is welcome to browse Appendices~\ref{app:General_ME} and \ref{app:Tunnel_Rates}, in which we highlight the breakdown of the semiclassical approach. To make the latter more accessible to the reader, we make an effort to explain key mathematical properties used during the derivation.
In Section~\ref{Regimes}, we look at the results from the point of view of equivalent-circuit simulation, and focus on measurable quantities of experimental relevance, predicting novel regimes and proposing experiments within reach of the current state of the art.

\subsection{Gate Current and Equivalent Quantum Admittance}

\begin{figure}[h!]
  \includegraphics[width = \linewidth]{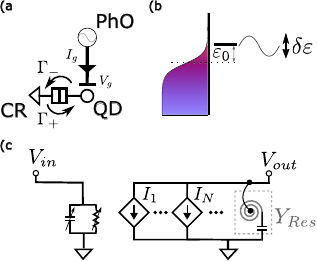}
  \caption{Schematics of the driven single electron box as its low-frequency circuit equivalent (a) and energy level diagram (b). The AC voltage source is seen as a Photonic Oscillator (PhO), and ground is seen as a Charge Reservoir (CR), connected to the QD via a tunnel barrier. Panel (c) shows an effective high-frequency model of the system, thought of as a parallel combination of the variable tunnelling capacitance and Sisyphus resistance as seen from the input and as a combination of parallel of voltage-controlled current sources at all the harmonics of the drive as seen from the output. In the schematic, a resonant circuit is added as a spectroscopic tool to investigate specific harmonics.}
  \label{fig:fig1}
\end{figure}

A SEB consists of a single QD connected via a tunnel barrier to a CR (at temperature $T$) and whose electrochemical potential can be capacitively changed via a gate, see Fig.~\ref{fig:fig1}a. 
We consider the CR to be connected to ground. In this work, we focus on the case in which a sinusoidally varying gate potential drives the SEB, 
\begin{equation}
  V_g(t) = V_0 + \delta V_g \cos{\omega t},
\end{equation}

\noindent and calculate the resulting gate current $I_g(t)$ due to CR-QD charge tunnelling events to understand the functional relationship between $V_g$ and $I_g$ introduced by the SEB.

More specifically, the induced gate charge due to a charge tunnelling event reads
\begin{equation}
  Q_g (t) = \alpha e P(t),
\end{equation}
\noindent
where $P(t)$ is the time-dependent probability of occupation and $\alpha$ is the lever arm that couples the gate to the QD. 
For legibility, in this Section, we will not distinguish between driving and collection lever arms (see Appendix~\ref{lineshape}). 
The gate current can be expressed as 
\begin{equation}
  I_g (t) = \alpha e \frac{d}{dt} P(t),
  \label{eq:GateI}
\end{equation}
\noindent
which is the quantity we shall derive in this work.
Previous works have focused on the small-signal regime where the SEB can be replaced by an equivalent admittance linking $I_g$ and $V_g$~\cite{Vigneau_2023, Esterli2019, Mizuta2017,Derakhshan_2020}. However, such a picture fails outside that limit due to the inherent nonlinearities of the quantum dynamics in the SEB. 
In this work, we will prove that the concept of an equivalent admittance can be extended to arbitrarily sized signals, as the gate current can be expressed as a Fourier series containing higher-order harmonics as
\begin{equation}
  I_g(t) = \sum_N I_N(t) = \frac{\delta V_g}{2} \sum_N Y_N e^{i N \omega t} + c.c.
  \label{eq:IN_def}
\end{equation}
\noindent
where the equivalent quantum admittance at the $N$-th harmonic is
\begin{equation}
  Y_N = \frac{\int_0^{\frac{2 \pi}{\omega}} e^{i N \omega t} I_g(t) dt}{\int_0^{\frac{2 \pi}{\omega}} e^{i \omega t} V_g(t) dt}.
  \label{eq:YN_def}
\end{equation}

Equation~\ref{eq:YN_def} becomes of particular physical interest when the SEB is coupled with an electrical resonator, for example, in a reflectometry setup, \cite{Vigneau_2023,oakes_fast_PRX,oakes2022quantum}, allowing the exploration of the different Fourier terms. An equivalent circuit is best thought of as in Fig.~\ref{fig:fig1}c, with the SEB appearing as a parallel combination of a nonlinear resistor and capacitor to the \textit{input} of the setup, while its effect on the \textit{output} is best modelled as a parallel of Voltage-Controlled Current Sources (VCCS) associated to the aforementioned Fourier terms, whose output current is then filtered by the resonator (Appendix~\ref{lineshape}). 
From a quantum perspective, the role of the resonator can be thought of as an electrodynamic cavity coupled to the SEB, whose role is simultaneously to amplify the resonant signals and select the desired harmonic.

In Appendix~\ref{lineshape}, we show how, if the $N$-th harmonic is the only one in the bandwidth of the resonator, one can approximate, in the usual regime where $|Y_{Res}| \gg |Y_N|$,
\begin{equation}
  V_{out}^N = \frac{Y_N}{Y_{Res}} V_{in}
\end{equation}
\noindent
where $V_{out}^N(t) = \Re\left[V_{out} e^{i N \omega t}\right]$ and $V_{in}(t) = \Re\left[V_{in} e^{i \omega t}\right]$ and $Y_{Res}$ is the resonator admittance.
Therefore, one only needs to determine the equivalent admittance of the system at the relevant harmonic to determine its reflection and transmission coefficients through a cavity. We emphasise that this result and  the methodology developed are \textit{general} for any QD system.

Finally, we shall notice from \eqr{eq:GateI} that $Y_N$ only depends on the probability of the QD being occupied. In Section~\ref{General_ME} we show how, the time evolution of the SEB can be generally expressed in a master equation of the form 
\begin{equation}
  \frac{d}{dt} P = - \Gamma P + \Gamma_- (t),
\end{equation}
\noindent
where $\Gamma$ is the total QD-CR tunnel rate (independent of time because of conservation of charge) and $\Gamma_- (t)$ is the tunnel rate \textit{out} of the QD. Thus, the admittance solely depends on the quantity $\Gamma_- (t)$ (which, as well as $\Gamma$, has units of Hz throughout this work). 
In this work, we shall present a self-consistent fully quantum way of deriving such a rate, which, when compared with the semiclassical result used thus far in the literature, will shed new light on the high-frequency behaviour of the SEB. 

In particular, in Section~\ref{Tunnel_Rates} we derive a general expression for $Y_N$, and in Section~\ref{Regimes} we identify different regimes differentiated by the physical process dominating the charge dynamics. While we present a complete and thorough discussion of the various regimes, for the benefit of the reader Tables~\ref{tab:regimes} and \ref{tab:regimes_N} present a short summary of the result derived in this work. A short glossary of the relevant symbols can be found in Tab.~\ref{tab:glossary}.
In particular, we emphasise the response of the SEB at its fundamental frequency $N=1$ (Tab.~\ref{tab:regimes}), as this is the result of most experimental interest.

\begin{table}[hb]
  \small{
  \begin{tabular}{p{0.3 \linewidth} p{0.65 \linewidth}}
    \toprule
    \multicolumn{2}{c}{Glossary of Symbols}\\
    \midrule
    $\varepsilon_0$ & Average QD-CR electrochemical potential detuning \\
    $\delta \varepsilon$ & Amplitude of QD-CR detuning variation \\
    $\omega$ & Frequency of the PhO\\
    $\Gamma$ & QD-CR tunnel rate \\
    $T$ & CR charge temperature \\
    $\kappa$ & PhO photon loss rate\\
    $g$ & QD-PhO coherent coupling \\
    $\gamma = \frac{\kappa \omega^2}{(2 g / \hbar)^2 + \kappa^2}$ & Photon Loss Broadening Rate\\
    $\alpha$ & Gate lever arm \\
    $J_n$ & $n$-th Bessel Function of the first kind\\
    $\psi_0$ & Digamma function\\
    $\psi_1$ & Trigamma function\\
    \bottomrule
  \end{tabular}}
\caption{Glossary of the symbols present in the effective SEB admittance (Tabs.~\ref{tab:regimes} and \ref{tab:regimes_N}) and their physical meaning.}
\label{tab:glossary}
\end{table}

\begin{table*}[htb!]
  \begin{center}
    \begin{tabular}{||>{\centering\arraybackslash}m{0.25 \textwidth}|>{\centering\arraybackslash}m{0.7 \textwidth}||}
      \hline
      \textbf{Regimes}    & \textbf{Fundamental Admittance ($Y_1$)}\\
      \hline
    \end{tabular}
    \begin{tabular}{||>{\centering\arraybackslash \hspace{0pt}
        \vfill}m{0.25 \textwidth}<{\hspace{0pt}
        \vfill}|>{\centering\arraybackslash \hspace{0pt}
        \vfill}m{0.7 \textwidth}<{\hspace{0pt} \vspace{5pt}
        \vfill }||}
      \hline
      Thermal Broadening (TB) \newline $k_B T \gg h \Gamma, \hbar \omega, \delta \varepsilon$ & $\displaystyle \frac{(\alpha e)^2}{2k_B T} \frac{\Gamma \omega}{\omega - i \Gamma} \cosh^{-2} {\parens{\frac{\varepsilon_0}{2 k_B T}}}$ \\
      \hline
      Lifetime Broadening (LB) \newline $h \Gamma \gtrsim k_B T \gg \hbar \omega, \delta \varepsilon$ & $\displaystyle \frac{(\alpha e)^2}{2 \pi^2 k_B T} \frac{\Gamma \omega}{\omega - i \Gamma} \Re\left[\psi_1\left(\frac{1}{2} + i \frac{\varepsilon_0}{2 \pi k_B T}+ \frac{h \Gamma}{2 \pi k_B T}\right)\right]$\\
      \hline
      Floquet Broadening (FB) \newline $\hbar \omega \gtrsim k_B T, h \Gamma \gg \delta \varepsilon$ & 
      $\displaystyle \begin{aligned} \frac{(\alpha e)^2}{2\pi \hbar} \frac{\Gamma}{\omega - i \Gamma} \Im\bigg[&\psi_0\left(\frac{1}{2} + i \frac{\varepsilon_0 +\hbar \omega}{2 \pi k_B T}+ \frac{h \Gamma}{2 \pi k_B T}\right) - \\ &\psi_0\left(\frac{1}{2} + i \frac{\varepsilon_0 -\hbar \omega}{2 \pi k_B T}+ \frac{h \Gamma}{2 \pi k_B T}\right)\bigg] \end{aligned} $ \\
      \hline
      Power Broadening (PB) \newline $\displaystyle \delta \varepsilon \gg k_B T , h \Gamma, \hbar \omega$ &
      $\displaystyle \begin{aligned} \frac{2 \hbar}{\pi} \left(\frac{\alpha e}{\delta \varepsilon}\right)^2\frac{\Gamma \omega^2}{\omega - i \Gamma} \sum_{m= - \infty}^{+ \infty} &m J_m \left( \frac{\delta \varepsilon}{\hbar \omega}\right) ^2
       \\ &\Im\left[\psi_0\left(\frac{1}{2} + i \frac{\varepsilon_0 + m \hbar \omega}{2 \pi k_B T}+ \frac{h \Gamma}{2 \pi k_B T}\right)\right] \end{aligned}$\\
      \hline
      Photon Loss Broadening (PLB) \newline $\displaystyle \delta \varepsilon \gtrsim \hbar \omega \sqrt{\Gamma/\gamma}$ \newline $\displaystyle \kappa \ll \Gamma + \gamma (\delta \varepsilon / \hbar \omega)^2$ &
      $\displaystyle \begin{aligned}\frac{2 \hbar}{\pi} \left(\frac{\alpha e}{\delta \varepsilon}\right)^2\frac{\Gamma \omega^2}{\omega - i \Gamma} \sum_{m= - \infty}^{+ \infty} &m J_m \left( \frac{\delta \varepsilon}{\hbar \omega}\right) ^2
      \\ &\Im\left[\psi_0\left(\frac{1}{2} + i\frac{\varepsilon_0 + m\hbar \omega }{2 \pi k_B T} + \frac{\Gamma + \gamma (\delta \varepsilon / \hbar \omega)^2}{2 \pi k_B T}\right)\right] \end{aligned}$ \\
      \hline
    \end{tabular}
  \end{center}
\caption{Quantum admittance at the fundamental of the driving frequency $Y_1$ of the SEB in the various regimes presented in this work: TB (Sec.\ref{thermal_broadening}), LB (Sec.\ref{Lifetime_Broad}), FB (Sec.\ref{Polaron_Broad}), PB (Sec.\ref{Power_Broad}), and PLB (Sec.\ref{Measurement_broad}). A glossary of the symbols and their physical interpretation can be found in Tab.~\ref{tab:glossary}. $\Re$/$\Im$ indicate the real and imaginary parts of the complex argument.}
\label{tab:regimes}
\end{table*}

\begin{table*}[htb!]
  \begin{center}
    \begin{tabular}{||>{\centering\arraybackslash}m{0.25 \textwidth}|>{\centering\arraybackslash}m{0.7 \textwidth}||}
      \hline
      \textbf{Regime}  & \textbf{$N$-th Harmonic Admittance ($Y_N$)} \\
      \hline
    \end{tabular}
    \begin{tabular}{||>{\centering\arraybackslash \hspace{0pt}
      \vfill}m{0.25 \textwidth}<{\hspace{0pt}
      \vfill}|>{\centering\arraybackslash \hspace{0pt}
      \vfill}m{0.7 \textwidth}<{\hspace{0pt} \vspace{5pt}
      \vfill }||}
      \hline
      Thermal Broadening (TB) \newline $k_B T \gg h \Gamma, \hbar \omega, \delta \varepsilon$ & 
      $\displaystyle (\alpha e)^2  \frac{\Gamma N \omega}{N \omega - i \Gamma} \frac{\partial^N}{\partial \varepsilon_0^N} \parens{1+e^{\varepsilon_0/k_B T}}^{-1}$\\
      \hline
      Lifetime Broadening (LB) \newline $h \Gamma \gtrsim k_B T \gg \hbar \omega, \delta \varepsilon$ &
      $\displaystyle \frac{(\alpha e)^2}{\pi} \frac{\Gamma N \omega}{N \omega - i \Gamma} \frac{\partial^N}{\partial \varepsilon_0^N} \Im\left[\psi_0\left(\frac{1}{2} + i \frac{\varepsilon_0}{2 \pi k_B T}+ \frac{h \Gamma}{2 \pi k_B T}\right)\right]$\\
      \hline
      \vspace{10pt} \vfill Floquet Broadening (FB) \newline $\hbar \omega \gtrsim k_B T, h \Gamma \gg \delta \varepsilon$ \newline&  
      \multirow{2}{=}{ \vspace{4pt} \vfill
      $\displaystyle \quad \begin{aligned} \frac{(\alpha e)^2}{\pi \delta \varepsilon} \frac{N \Gamma \omega}{N\omega - i \Gamma} \sum_{m= - \infty}^{+ \infty} J_m \left( \frac{\delta \varepsilon}{\hbar \omega}\right) 
      \bigg(
        &J_{m+N} \left( \frac{\delta \varepsilon}{\hbar \omega}\right) +
        J_{m-N} \left( \frac{\delta \varepsilon}{\hbar \omega}\right)
      \bigg)  \\
      &  \Im\left[\psi_0\left(\frac{1}{2} + i \frac{\varepsilon_0 + m \hbar \omega}{2 \pi k_B T}+ \frac{h \Gamma}{2 \pi k_B T}\right)\right] \end{aligned} $
       }\\
      \cline{0-0}
      \vspace{10pt} \vfill Power Broadening (PB) \newline $\delta \varepsilon \gg k_B T , h \Gamma, \hbar \omega$  \newline & \\
      \hline
      Photon Loss Broadening (PLB) \newline $\displaystyle \delta \varepsilon \gtrsim \hbar \omega \sqrt{\Gamma/\gamma}$ \newline $\displaystyle \kappa \ll \Gamma + \gamma (\delta \varepsilon / \hbar \omega)^2$ \vspace{4pt}
      \vfill &
      $ \displaystyle \quad \begin{aligned} \frac{(\alpha e)^2}{\pi \delta \varepsilon} \frac{N \Gamma \omega}{N\omega - i \Gamma} \sum_{m= - \infty}^{+ \infty} J_m & \left( \frac{\delta \varepsilon}{\hbar \omega}\right) 
      \bigg( 
        J_{m+N} \left( \frac{\delta \varepsilon}{\hbar \omega}\right)+
        J_{m-N} \left( \frac{\delta \varepsilon}{\hbar \omega}\right)
      \bigg)  \\
      & \Im\left[\psi_0\left(\frac{1}{2} + i\frac{\varepsilon_0 + m\hbar \omega }{2 \pi k_B T} + \frac{\Gamma + \gamma (\delta \varepsilon / \hbar \omega)^2}{2 \pi k_B T}\right)\right]& \end{aligned}$\\
      \hline
    \end{tabular}
\end{center}
\caption{Quantum admittance at the $N$-th harmonic of the driving frequency $Y_N$ of the SEB in the various regimes presented in this work: TB (Sec.\ref{thermal_broadening}), LB (Sec.\ref{Lifetime_Broad}), FB (Sec.\ref{Polaron_Broad}), PB (Sec.\ref{Power_Broad}), and PLB (Sec.\ref{Measurement_broad}). A glossary of the symbols and their physical interpretation can be found in Tab.~\ref{tab:glossary}. $\Re$/$\Im$ indicate the real and imaginary parts of the complex argument.}
\label{tab:regimes_N}
\end{table*}

\subsection{Reflectometry as Polaron Dynamics}
\label{Reflecto_as_Polaron}

While the main result of this work is the derivation in closed form of the effective SEB admittance $Y_N$, a prerequisite was to derive a self-consistent fully quantum formalism for an open quantum system subject to sinusoidal driving. The results and mathematical toolbox assembled in this process lay the foundation to go beyond semiclassical master equations in the description of equivalent electronic circuits. In particular, describing a driven QD as a polaron sheds new light on the concept of gate current and the process of reflectometry, also serving as a cautionary tale when semiclassical methods venture beyond the limits imposed by the Heisenberg Uncertainty Principle.

The quantum description of the SEB begins by considering the semiclassical drive as the manifestation of (weak) coupling with a (coherent) Photonic Oscillator (PhO), which is held by a source in a coherent state. 
We can now begin by describing the SEB in a second-quantization formalism, as outlined in Section~\ref{Hamiltonian_def}, by representing the dynamics via the annihilation of the QD, CR, and PhO and their respective adjoint operators. 

Semiclassically, we would group the systems in Fig.~\ref{fig:fig1}a \textit{horizontally}, i.e., consider first the QD as coupled with a CR, which preserves the conservation of charge during the time evolution. However, this leads our intuition to think of the SEB as a \textit{completely incoherent} system, where the effect of the PhO is merely to \textit{drive} stochastic tunnelling events in and out of the QD.  
Formally, as we show in Section~\ref{Tunnel_Rates}, this relies on the Instantaneous Eigenvalues Approximation \cite{Yamaguchi_Yuge_Ogawa_2017}, which consists of a secular approximation of the quantum dynamics, where the driven QD acts as a single level whose energy periodically oscillates in time. Such an adiabatic treatment of the problem forgoes the quantum nature of the dynamics.
In particular, while the final master equation will be Markovian and thus time-local, the expression to derive the tunnel rates necessarily contains a time convolution \cite{manzano_short_2020}, preventing a simple extension of the adiabatic result in the time-dependent case \cite{Mori_2023}. 

We shall see, however, how the Lindblad formalism drastically simplifies when we group Fig.~\ref{fig:fig1}a \textit{vertically} and consider a mixed electron-photon state (i.e., a polaron) whose fermionic part stochastically interacts with the CR. 
Most interestingly, in Section~\ref{self_ME}, we show how the CR-polaron interaction reads
\begin{equation}
  H_{CR-Pol} \propto c^{\dagger} D d + h.c.,
  \label{eq:polaron_interaction}
\end{equation}
\noindent
where $c$ and $d$ are the annihilation operators of the CR and QD, respectively.
\begin{equation}
  D = \exp\parens{- \frac{g}{\hbar \omega} \parens{a^\dagger - a}}
  \label{eq:displacement_def}
\end{equation}
\noindent
is known as a displacement operator of the PhO, where $a$ is the annihilation operator of the PhO, $\hbar \omega$ is the photon energy, and $g$ is the coherent QD-PhO coupling.

Equation~(\ref{eq:polaron_interaction}), therefore, dictates that to add (remove) an electron in the QD, one must also \textit{displace} the PhO.
Fundamentally, this shows that in a driven QD, a tunnelling event can only occur through the creation or annihilation of a polaron.
For the benefit of the reader, we recall that the polaron frame is defined by the canonical transformation 
\begin{equation}
  e^S = \exp \left(- \frac{g}{\hbar \omega} \parens{a^\dagger - a} d^\dagger d \right)
  \label{eq:polaron_frame_def}
\end{equation}
\noindent i.e., a photonic system that is displaced \textit{selectively} based on the state of a (fermionic) system, see Section~\ref{self_ME}.
Equation~(\ref{eq:polaron_interaction}) thus gives a direct physical interpretation of the concept of gate current. In the Lindblad model, we assume the PhO to remain unperturbed by the (weak) coupling to the QD. The gate current is a manifestation of those \textit{extra} photons that change state as a back-action of charge tunnelling event, and which manifest semiclassically as an alternating current (AC).

This formalism allows us to go beyond the limits of the semiclassical approach.
In particular, we can describe the effects of a high-frequency drive or a short lifetime of the charge in the QD, both cases in which the semiclassical picture breaks the time-energy Uncertainty Principle. 
By describing reflectometry as polaron dynamics, we shall see that the Heisenberg principle is recovered in both limits, which will be reflected in the reflectometry lineshapes at the fundamental $Y_1$.
Notably, we stress that this \textit{must} be the case. 
Formally, we can think of the limit of a zero temperature CR as a \textit{spectroscopic} probe of the effective density of states of the SEB. In this case, $Y_1$ takes the exact form of the \textit{spectrum} of the system \cite{QuantumNoise}, and thus, in a rigorous quantum description, the Heisenberg uncertainty principle must be satisfied.
More physically, we note that $Y_1$ directly derives from the Fourier transform of a physical observable, i.e., the charge in the QD. Considering that the time-energy uncertainty principle derives from the relation between real and reciprocal space, we expect it to be conserved.

\section{Quantum Dynamics of The Single-Electron Box}
\label{SEB_dynamics}

In this section, we shall define the SEB in terms of its separate subsystem: a QD, a CR, and a PhO. We will also show that any master equation in Lindblad form can be written without loss of generality in terms of a single Ordinary Differential Equation for the probability of occupation of the QD as a function of time, thanks to well-known normalization properties of the density matrix of a system.

\subsection{System Hamiltonian}
\label{Hamiltonian_def}

Regarding its separate subsystems, the complete SEB Hamiltonian reads

\begin{equation}
  H = H_{QD} + H_{CR} + H_{PhO} + H_{DR} + H_{DP},
  \label{eq:H_tot}
\end{equation}
\noindent
where $H_{QD}$, $H_{CR}$, and $H_{PhO}$ represent the unperturbed Hamiltonians of the QD, CR and PhO, respectively. While $H_{DR}$ and $H_{DP}$ describe the QD-CR and QD-PhO coupling, respectively. In particular, we take the zero of energy at the Fermi level of the CR and consider Hamiltonians of the form
\begin{align}
  &H_{QD} = \varepsilon_0~ d^\dagger d\\
  &H_{R} = \sum_\epsilon ~\epsilon ~c_\epsilon^\dagger c_\epsilon\\
  &H_{PhO} = \hbar \omega ~ a^\dagger a,
\end{align}

\noindent with $d$, $c$, and $a$ representing the destruction operators of the QD, CR, and PhO, respectively. The parameter $\varepsilon_0$ describes the static energy detuning of the single QD level (Fig.~\ref{fig:fig1}a), and $\epsilon$ is the energy of the (potentially discrete) CR levels. We assume the microwave radiation to be monochromatic at the single frequency $\omega$. 
For ease of future notation, we can define 
\begin{equation}
  H_0 = H_{QD} + H_{R} + H_{PhO}
\end{equation}
\noindent
as the Hamiltonian describing the free evolution of the (uncoupled) quantum systems. Those are then coupled by interaction Hamiltonians, which in the second quantization formalism are \cite{Sowa_Mol_Briggs_Gauger_2018}
\begin{align}
  &H_{DR} = \sum_\epsilon ~ V_\epsilon c_\epsilon d^\dagger + V_\epsilon^* c_\epsilon^\dagger d
  \label{eq:CR_QD_coupl}\\
  &H_{DP} = g(a + a^\dagger) d^\dagger d,
\end{align}
\noindent
where we have considered the experimentally relevant case of longitudinal QD-PhO coupling. 
The parameter $g$ is the coherent QD-PhO coupling, and $V_\epsilon$ is the CR-QD coupling at energy $\epsilon$.
For the sake of mathematical simplicity, and because it is not usually experimentally relevant, we neglect any direct PhO-CR interaction. 
We point out that, in the semiclassical picture, this is equivalent to neglecting geometrical contributions to the admittance seen by the gate \cite{Mizuta2017,Esterli2019}.

\subsection{Master Equation and Charge Dynamics}
\label{General_ME}

Equation~\eqref{eq:H_tot} describes the complete quantum dynamics of the QD-CR-PhO system. In this work, we are interested in the QD charge dynamics alone. Therefore, in the subsequent sections, we shall employ the Lindblad formalism to write a master equation that only describes the QD degree of freedom by appropriately tracing over the CR and PhO. The final Lindblad master equation (LME) will take the standard form 

\begin{equation}
  \hbar \frac{d}{dt} \rho(t) = -i [H_0(t), \rho(t)] + \sum_{i = +,-} h \Gamma_i(t) \mathcal{D}[L_i],
  \label{eq:LME_main}
  \end{equation}
\noindent 
where
\begin{equation}
      \mathcal{D}(L_i)[\rho(t)] = L_i^\dagger \rho(t) L_i - \frac{1}{2} \left\{ L_i^\dagger L_i, \rho(t) \right\}
  \end{equation}
\noindent 
are the dissipation superoperators that account for the non-unitary dynamics caused by the coupling to the CR. The latter is described by the \textit{jump} operators 
\begin{equation}
\begin{aligned}
  &L_+ = d^\dagger\\
  &L_- = d
\end{aligned}
\label{eq:jump_op}
\end{equation}
\noindent
linked to the corresponding tunnel rates $\Gamma_{\pm}(t)$, which we shall derive in the subsequent sections. 

We can simplify the problem further by formulating it as a two-level system, where the QD is either occupied ($\ket{o}$) or empty ($\ket{e}$). A closer look at the jump operators in \eqr{eq:jump_op} leads to identifying $\Gamma_{\pm}(t)$ as the probability per unit time of an electron entering (exiting) the QD from the CR. 
Therefore, a simple argument based on the Fermi Golden Rule leads to the rate of jumping in (out) of the QD only depending on the number of occupied (empty) states in the CR. Consequently, 
\begin{equation}
  \Gamma_+ (t) + \Gamma_- (t) = \Gamma,
  \label{eq:Fermionc_Character}
\end{equation}
\noindent
with $\Gamma$ being a constant representing the total charge tunnel rate~\cite{StochasticMethods, QuantumNoise}.
Moreover, the previous interpretation of the QD as a \textit{two}-level system allows us to characterize the quantum state of the system with the density matrix
\begin{equation}
  \rho = \begin{pmatrix}
    P & C\\
    C^* & 1-P
  \end{pmatrix},
  \label{eq:rho_def}
\end{equation}
\noindent
where $P(t)$ is the probability of occupation of the QD and $C = \textnormal{tr}\{|e\rangle \langle o| \rho\}$ quantifies the degree of coherent superposition between the two states. In \eqr{eq:rho_def} we already enforce the normalization $\textnormal{tr}\{\rho\}=1$.

In Appendix~\ref{app:General_ME} we show how, even if the system starts with nonzero coherence, $C$ vanishes exponentially in time. Therefore, after a (short) transient of timescale $1/\Gamma$, the SEB becomes a statistical mixture of the two states, completely described by the occupation of the QD.
Moreover, Appendix~\ref{app:General_ME} also shows how \eqr{eq:Fermionc_Character} mandates that the time evolution of $P(t)$ always follows the equation
  \begin{equation} 
    \frac{d}{dt} P + \Gamma P = \Gamma_- (t),
    \label{eq:pop_ME}
  \end{equation}
  \noindent
which trivially descends from the normalization of the density matrix.
We can immediately write the steady-state solution in the form
  \begin{equation}
    P(t) = e^{-\Gamma t} \int_{-\infty}^t e^{\Gamma \xi} \Gamma_-(\xi) ~ d \xi.
    \label{eq:P_sol}
  \end{equation}
\indent Considering the time symmetries of the problem, we expect (as we shall confirm in Section~\ref{Tunnel_Rates}) that $\Gamma_\pm(t)$ will be a periodic function with the same period $2 \pi/\omega$ as the PhO. Therefore, \eqr{eq:P_sol} can be rewritten as \cite{oakes2022quantum}
  \begin{equation}
    P(t) = \frac{\omega}{2 \pi} \sum_N \frac{e^{i N \omega t}}{\Gamma + i N \omega} \int_0^{\frac{2 \pi}{\omega}} e^{i N \omega t} \Gamma_-(t) ~~dt + c.c.
  \end{equation} 
Recalling the definition of gate current and quantum admittance in Eqs.~(\ref{eq:GateI}) and (\ref{eq:YN_def}), we can therefore write 
\begin{equation}
  \begin{aligned}
  Y_N &= i \frac{(e \alpha)^2}{\delta \varepsilon} \frac{N\omega}{\Gamma + i N \omega} \int_0^{\frac{2 \pi}{\omega}} e^{i N \omega t} \Gamma_-(t) ~~dt= \\ 
  &= \frac{\mathcal{C}_N}{\delta \varepsilon} \int_0^{\frac{2 \pi}{\omega}} e^{i N \omega t} \frac{\Gamma_-(t)}{\Gamma} ~~dt,
  \end{aligned}
  \label{eq:YN_general}
\end{equation}
\noindent
where, for conciseness, we have defined the coefficient 
\begin{equation}
  \mathcal{C}_N = i (e \alpha)^2 \frac{N\omega \Gamma}{\Gamma + i N \omega},
  \label{eq:CN}
\end{equation}
which accounts for the semiclassical high-pass filtering effect of the capacitive coupling of the collection gate.
Notably, in a fermionic system the level crossing between $\ket{e}$ and $\ket{o}$ cannot be avoided as the number of particles in the system is changing, and tunnelling is stochastic in nature \cite{cochrane2022intrinsic}. Therefore, $Y_N$ doesn't contain any capacitance arising from the curvature of the energy levels (sometimes referred to in the literature as \textit{quantum} capacitance) \cite{Ashoori_1992, Ashoori_1993, Peri_2023,Persson_Wilson_Sandberg_Johansson_Delsing_2010,Ciccarelli_2011,Vigneau_2023}, as highlighted throughout the text and in Appendix~\ref{sect_GC}.
From \eqr{eq:YN_general}, we arrive at the key insight that the equivalent admittance of the SEB is uniquely determined by the functional form of $\Gamma_-(t)$. In the next section, we investigate the the derivation of the tunnel rates in different approximations and levels of theoretical complexity: the semiclassical limit (Section \ref{Semiclassical_gamma}) and with a fully quantum self-consistent approach (Section \ref{self_ME}).

\section{Tunnel Rates and Effective Admittance}
\label{Tunnel_Rates}

In the previous section, we showed how the properties of the charge dynamics and the effective admittance of the SEB solely depend on the functional form of the tunnel rates. Here, we present the derivation of $\Gamma_-(t)$ in different limits of theoretical complexity: the semiclassical limit (Section \ref{Semiclassical_gamma}) and with a fully quantum self-consistent approach (Section \ref{self_ME}). 
In both cases, we can simplify the derivation making use of the general expression \cite{Yan_1998,Sowa_Mol_Briggs_Gauger_2018}
\begin{equation}
\begin{aligned}
  \hbar \frac{d}{dt} \rho& =  -i \left[ H_0(t), \rho(t) \right] - \\
  &-\int_0^{+ \infty} d\tau \langle \mathcal{L}'(t) \mathcal{G}(t, \tau) \mathcal{L}'(\tau)\mathcal{G}^\dagger(t, \tau) \rangle \rho(t),
\end{aligned}
  \label{eq:ME_gen}
\end{equation}
\noindent
which we shall employ to determine the tunnel rates. In \eqr{eq:ME_gen}, $\mathcal{G}(t, \tau)$ is the propagator of the (unperturbed) QD dynamics, while $\mathcal{L}'(t)$ is the superoperator describing the perturbation due to the coupling between the QD and the CR and PhO. The notation $\langle \cdot \rangle$ indicates the partial trace over degrees of freedom not of the QD.
The aim of this section is to present physical insights into the results of this work, with particular care to highlight the breakdown of the semiclassical approximation and the importance of quantum theory in describing the SEB. For the benefit of the reader interested in the mathematical methods employed, we report a full derivation of the tunnel rates in Appendix~\ref{app:Tunnel_Rates}.

\subsection{Semiclassical Master Equation}
\label{Semiclassical_gamma}

A simple method to describe the longitudinal QD-PhO coupling is to write
\begin{equation}
  H_0(t) = \braket{H_{QD} + H_{DP}}_{PhO} = \parens{\varepsilon_0 + \delta \varepsilon \cos{\omega t}} d^\dagger d,
  \label{eq:H_sc}
\end{equation}
\noindent
where $\braket{\cdot}_{PhO}$ represents the partial trace over the PhO, and we have used the fact that 
\begin{equation}
  \braket{g\parens{a^\dagger + a}}_{PhO} = \delta \varepsilon \cos{\omega t}.
  \label{eq:field}
\end{equation}

We will derive this more thoroughly in the subsequent section. It stems from the semiclassical insight that $\parens{a^\dagger + a}$ is proportional to the electric-field operator, which we expect to be monochromatic. We can extend this parallel more deeply by observing from Eq.~\eqref{eq:H_sc} that $\delta \varepsilon$ represents the amplitude of the energy-detuning oscillation. Therefore, we can rederive the conventional result by identifying 
\begin{equation}
  \delta \varepsilon = \alpha e \delta V_g,
\end{equation}
\noindent
where $\delta V_g$ is the peak-to-peak voltage amplitude applied to the QD gate and $\alpha$ the correspondent gate lever arm (Fig.~\ref{fig:fig1}a) \cite{Peri_2023}. Therefore, the longitudinal coupling is the quantum equivalent of a sinusoidally oscillating detuning 
\begin{equation}
\varepsilon(t) = \varepsilon_0 + \delta \varepsilon \cos{\omega t}.
\label{eq:detuning_semicl}
\end{equation}

We shall note that Eq.~\eqref{eq:H_sc} is formally obtained by considering the PhO to first order in the Lindblad theory and describing the system as an interacting QD-CR pair following the time-dependent Hamiltonian
\begin{equation}
  H_{SC} = H_0(t) + H_R + H_{DR}.
  \label{eq:H_sc2}
\end{equation}
This is the semiclassical description of an SEB considered thus far in the literature \cite{Mizuta2017,Esterli2019}.
Notably, in this picture, Eqs.~(\ref{eq:field}) and (\ref{eq:detuning_semicl}) are equivalent to a simple capacitive model, where $\varepsilon(t) = \alpha e V_g(t)$ \cite{Vigneau_2023,Esterli2019}.

In Appendix~\ref{app:Semiclassical_gamma} we show how, in this semiclassical approximation and in the wide-band limit of the CR \cite{Sowa_Lambert_Seideman_Gauger_2020}, we obtain the well-known result \cite{cochrane2022intrinsic,Cochrane2022,oakes_fast_PRX,oakes2022quantum,Vigneau_2023}
\begin{eqnarray}
  \Gamma_{\pm} (t) = \Gamma f(\mp \varepsilon(t)),
  \label{eq:sc_rates}
\end{eqnarray}
\noindent
where 
\begin{equation}
  f(\epsilon) = \frac{1}{e^{\epsilon/k_B T} + 1}
\end{equation}
is the Fermi-Dirac distribution.

We shall stress how, semiclassically, consistently with the name instantaneous eigenvalues approximation, the master equation only considers the reservoir states at the instantaneous energy of the QD, thus treating the level as a Dirac delta in energy space and neglecting any broadening that may occur because of the finite coupling to the CR. 
Moreover, it highlights the \textit{secular} (adiabatic) nature of the approximation. Therefore, it fails to consider that the dynamics generated by a periodically driven Hamiltonian are far richer than a simple oscillation of the QD energy.
Such an approximation is only valid when $\omega$ is much slower than any other timescale (i.e., $\Gamma$) \cite{Brasil_Fanchini_Napolitano_2013,Dann_Levy_Kosloff_2018,Yamaguchi_Yuge_Ogawa_2017,kohler_floquet-markovian_1997}, and thus we expect it to break down if the driving frequency is comparable with the other energies and timescales of the dynamics.
These concerns will be addressed by the self-consistent approach discussed in the next section.
In the following, we shall refer to \eqr{eq:pop_ME} with the rates defined in \eqr{eq:sc_rates} as the Semiclassical master equation (SME)~\cite{Gonzalez-Zalba2015}.

Finally, Eq.~(\ref{eq:sc_rates}) allows us to immediately compute the effective admittance arising from the semiclassical model, which reads
\begin{equation}
  Y^{SME}_N = 2\frac{\mathcal{C}_N}{\delta \varepsilon} \int_0^{\frac{2 \pi}{\omega}} e^{i N \omega t} f(\varepsilon_0 + \delta\varepsilon \cos{\omega t}) dt~.
\label{eq:Y_sc}
\end{equation}

\subsection{Self-Consistent Quantum Master Equation}
\label{self_ME}

In the previous section, we have seen how the main approximations of the semiclassical model stem from tracing out the quantum degrees of freedom of the PhO, and from neglecting any broadening of the density of states of the (metastable) level of the QD.
Here, we derive a fully quantum master equation, addressing both concerns by, respectively, treating the PhO quantum-mechanically and taking a self-consistent approach to the QD-CR interaction. 

The key insight that allows for a quantum treatment of the radiation is to consider the electron-photon interaction in the Lang-Firsov formalism and to perform a canonical transformation into the polaron frame of reference\cite{Jang_2022,Xu_Cao_2016,Wilner_Wang_Thoss_Rabani_2015}. This is achieved by considering the operator
\begin{equation}
S = - \frac{g}{\hbar \omega} \parens{a^\dagger - a} d^\dagger d
\end{equation}
\noindent
from \eqr{eq:polaron_frame_def} and studying the polaron-transformed Hamiltonian 
\begin{equation}
  \tilde{H} = e^{-S} H e^{S}.
\end{equation}

This is equivalent to applying the displacement operator in \eqr{eq:displacement_def}, which selectively displaces the PhO depending on the state of the QD. In Appendix~\ref{H_pol}, we show how, in the polaron frame,
\begin{equation}
  \begin{aligned}
   &e^{-S} \parens{H_{PhO}+ H_{QD} + H_{DP}} e^{S} =\\
  &\parens{\varepsilon_0 + \frac{g^2}{\hbar \omega}} d^\dagger d + \hbar \omega a^\dagger a,
  \end{aligned}
   \label{eq:Lamb_Shift}
\end{equation}
\noindent
where we can see the longitudinal coupling merely becomes a Lamb shift of the QD detuning, taking the form
\begin{equation}
  \tilde{H}_{QD}=\parens{\varepsilon_0 + \frac{g^2}{\hbar \omega}} d^\dagger d = \tilde{\varepsilon}_0 d^\dagger d.
\end{equation}

Moreover, in Appendix~\ref*{H_pol}, by employing the Baker-Campbell-Hausdorff (BCH) theorem, we find that \cite{Sowa_Lambert_Seideman_Gauger_2020,Sowa_Mol_Briggs_Gauger_2018}
\begin{equation}
  e^{-S} \parens{H_{DR}} e^{S} = \sum_\epsilon ~ V_\epsilon c_\epsilon D^\dagger d^\dagger + V_\epsilon^* c_\epsilon^\dagger D d\\
  \label{eq:HDR_pol}
\end{equation}
\noindent
as discussed in Section~\ref{Reflecto_as_Polaron}.

To include broadening of the QD level induced by the CR, we can recall that \eqr{eq:sc_rates} contains only on the \textit{instantaneous} energy of the QD. As detailed in Appendix~\ref{app:Semiclassical_gamma}, this originates from the use in \eqr{eq:ME_gen} of the \textit{free} propagator $\mathcal{G}(t, \tau)$, which only takes into account $H_{QD}$ \cite{Sowa_Mol_Briggs_Gauger_2018}. In Appendix~\ref{app:self_ME} we show how this can be remedied via a self-consistent Born approximation, which replaces the free propagator with a self-consistent superoperator $\mathcal{U}(t, \tau)$ that properly accounts for the metastable nature of an electron in the QD (Appendix~\ref{Digamma}) \cite{Sowa_Lambert_Seideman_Gauger_2020,Lee_2009}.
In Appendix~\ref{app:self_cons_prop}, we show how this approach
yields the intuitive result \cite{Sowa_Lambert_Seideman_Gauger_2020,Sowa_Mol_Briggs_Gauger_2018}
\begin{align}
  &\mathcal{U}(0, t)[d] = d e^{-i \tilde{\varepsilon}_0 t} e^{- \Gamma t} \\
  &\mathcal{U}(0, t)[d^\dagger] = d^\dagger e^{i \tilde{\varepsilon}_0 t} e^{- \Gamma t}~,
  \label{eq:self_consistent_time_evol}
\end{align}
\noindent
which has the simple effect of introducing a finite lifetime to the QD operators.

Carrying out the prescription of \eqr{eq:ME_gen} with these new addition leads to the tunnel rates (Appendix~\ref{app:self_cons_tunnel_rates})
\begin{equation}
  \begin{aligned}
  \Gamma_-(t) = \Gamma \sum_{m= - \infty}^{+ \infty} \sum_{n= - \infty}^{+ \infty} & J_n \left( \frac{\delta \varepsilon}{\hbar \omega}\right) J_m \left( \frac{\delta \varepsilon}{\hbar \omega}\right) \cdot\\ & \cdot \Re\left[ 
e^{ -i (m-n)\omega t}\mathcal{F}_m^{-}(\varepsilon_0)\right].
\end{aligned}
\label{eq:Gamma_scq_fourier_repl}
\end{equation}
This equation is the self-consistent quantum equivalent of the semiclassical \eqr{eq:sc_rates}. This result is general, and, for mathematical simplicity, only makes the approximation of weak QD-PhO coupling ($g \ll \hbar \omega$), as is common under the usual experimental conditions \cite{kohler_dispersive_2017,kohler_dispersive_2018,Benito_Mi_Taylor_Petta_Burkard_2017}. If these conditions are not met, then the PhO must be considered as a cavity which is part of the measured system \cite{Gu_Kohler_2023,Perez_2022,Koski_2018}. This only introduces a finite Lamb shift, and additional terms in the tunnel rates (Appendix~\ref{app:self_cons_tunnel_rates}).

Firstly, we note how Bessel functions of the first kind $J_n$ naturally appear in \eqr{eq:Gamma_scq_fourier_repl}, resembling previous results obtained in the non-adiabatic quantum master equation framework or Floquet-Lindblad formalisms \cite{Dann_Levy_Kosloff_2018,Yamaguchi_Yuge_Ogawa_2017,kohler_floquet-markovian_1997,Ikeda_Chinzei_Sato_2021,Mori_2023,Benito_Mi_Taylor_Petta_Burkard_2017}.
Moreover, similar functional forms have already been discussed in the literature  of superconducting structures (where the term SEB was coined) in the context of quantum mixing and photon detection \cite{Tien_Gordon_1963,Tucker_Feldman_1985}.

Moreover, in Appendix~\ref{app:self_ME}, we have defined
\begin{equation}
  \mathcal{F}_m^{-}(\varepsilon_0) = \frac{\Gamma}{\pi} \int_{-\infty}^{\infty} ~~~ \frac{f(\epsilon)}{\Gamma^2 + ((\epsilon - \varepsilon_0)/\hbar - m \omega)^2} 
~~~d\epsilon  
\label{eq:broad_fd_conv_main}
\end{equation}
\noindent
as the self-consistent distribution of the SEB, i.e., the convolution of the Fermi-Dirac distribution and a Lorentzian representing the effective density of states of the metastable QD level.

Within Appendix~\ref*{Digamma}, we showed how this (lifetime-) broadened Fermi-Dirac function can be rewritten analytically as 
\begin{equation}
  \mathcal{F}_m^{\pm}(\varepsilon_0)= \frac{1}{2} -\frac{1}{\pi} \Im\left[\psi_0\left(\frac{1}{2} + i \frac{\Theta^{\pm}_{m}}{2\pi} \right)\right],
  \label{eq:digamma}
\end{equation}
\noindent
where $\psi_0$ is Euler's digamma function and 
\begin{equation}
  \Theta^{\pm}_{m} = \pm \frac{\varepsilon_0 + m \hbar \omega}{\kt} - i \frac{h \Gamma}{\kt}.
    \label{eq:quasienergies}
\end{equation}
The numerator in \eqr{eq:quasienergies} can be interpreted as the (complex) quasi-energies of the metastable Floquet modes of the damped-driven SEB\cite{kohler_driven_2005} (as we shall see in Section~\ref{Floquet_Modes}).

We can now use the property that, thanks to the anticommutation relations of the fermionic CR operators, 
\begin{equation}
  \mathcal{F}_m^{-}(\varepsilon_0)= 1 - \mathcal{F}_{m}^{+}(\varepsilon_0)
\end{equation}
\noindent
to show that the property in \eqr{eq:Fermionc_Character} still holds and, thus, we obtain the Self-Consistent Quantum Master Equation (SCQME) of the form in \eqr{eq:pop_ME}.

We can obtain the effective admittance in ~\eqr{eq:YN_general} by selecting the correct Fourier component from \eqr{eq:Gamma_scq_fourier_repl}, yielding
\begin{equation}
  \begin{aligned}
  Y^{SCQME}_N &= \frac{\mathcal{C}_N}{\delta \varepsilon} \sum_{m= - \infty}^{+ \infty} J_m \left( \frac{\delta \varepsilon}{\hbar \omega}\right) \cdot \\
& \cdot \left(
J_{m+N} \left( \frac{\delta \varepsilon}{\hbar \omega}\right)+
J_{m-N} \left( \frac{\delta \varepsilon}{\hbar \omega}\right)
\right)
\mathcal{F}_m^{-}(\varepsilon_0).
\end{aligned}
\label{eq:Y_q}
\end{equation}

It is worth noting that in the case $N=1$, we can use the well-known identity
\begin{equation}
  J_{m+1}(x) + J_{m-1}(x) = 2\frac{m}{x} J_m(x)
  \label{eq:Bessel_centerpoint}
\end{equation}
to  write for the fundamental 
\begin{equation}
  \begin{aligned}
  &Y^{SCQME}_1 = 2 \frac{\hbar \omega}{\delta \varepsilon^2} \mathcal{C}_1 {\delta \varepsilon} \sum_{m= - \infty}^{+ \infty} m J_m^2 \left( \frac{\delta \varepsilon}{\hbar \omega}\right)
\mathcal{F}_m^{-}(\varepsilon_0) =\\ 
& = 2 \frac{\hbar \omega}{\delta \varepsilon} \mathcal{C}_1 \sum_{m= 1}^{+ \infty} m J_m^2 \left( \frac{\delta \varepsilon}{\hbar \omega}\right) \parens{\mathcal{F}_m^{-}(\varepsilon_0) - \mathcal{F}_{-m}^{-}(\varepsilon_0)},
\end{aligned}
\label{eq:Y_q_1}
\end{equation}
\noindent
where we have used the fact that $J_{-m}(x) = (-1)^m J_{m}(x)$.
Equations~(\ref{eq:Y_q}) and (\ref{eq:Y_q_1}) represent the major result of this work, as they are the most general expression of the SEB admittance (assuming weak photon coupling). However, because of their mathematical complexity, Section~\ref{Regimes} presents them in different regimes, where approximations can be made to simplify the expressions and thus highlight the effect of each parameter.

\subsection{A Floquet Interpretation}
\label{Floquet_Modes}

Before discussing the impact of the SCQME on the reflectometry lineshapes, we believe valuable to the reader to physically interpret \eqr{eq:Y_q} in terms of Floquet theory.

The first step consists of the solution of a sinusoidally driven single-level system whose Hamiltonian reads
\begin{equation}
  H = \varepsilon_0 + \delta \varepsilon \cos{\omega t}.
\end{equation}

As mentioned above, similar problems are well-known in the field of superconducting structures \cite{Tucker_Feldman_1985}, from which we can borrow an explicit solution for the Fourier decomposition of the single level, known as the Tien-Gordon model, which reads\cite{Tien_Gordon_1963,Platero_Aguado_2004}
\begin{equation}
  \ket{\phi (t)} = e^{i \varepsilon_0 t/\hbar} \sum_{m=-\infty}^{\infty} J_m\left(\frac{\de}{\hbar \omega}\right) e^{i m \omega t} \ket{\phi_m},
  \label{eq:Floquet_single}
\end{equation}
\noindent
where the Fourier components $\ket{\phi_m}$ traditionally take the name of Floquet modes. Notably, the complexity of materials and technologies has already been abstracted away in the charge dynamics of a driven discrete level, which makes \eqr{eq:Floquet_single} thus directly applicable also to QDs.
This result can be used to highlight key concepts of the dynamics of a driven quantum system. We can consider adding an integer offset $n$ to the index $m$, which would not change \eqr{eq:Floquet_single} apart from a redefinition of dummy indexes and the transformation $\varepsilon_0 \rightarrow \varepsilon_0 + n \hbar \omega$. Therefore, the energy of a periodically driven quantum system is defined up to an integer multiple of the photon energy. For this reason, they are traditionally named \textit{quasi}-energies. 
However, this simple picture invites us to think of a single driven level as a \textit{ladder} of energetically equally spaced Floquet modes, whose phase in time oscillates at the harmonics of the fundamental driving tone. This is sometimes referred to as the Fourier or Sambe decomposition of the state \cite{Rudner_Lindner_2020}.

This simple model, nonetheless, predicts an \textit{effective} density of states of the driven QD \cite{Tien_Gordon_1963} 
\begin{equation}
  \mathcal{D}_{QD} = \sum_{m=-\infty}^{+\infty} J_m^2\left(\frac{\de}{\hbar \omega}\right)   \frac{\Gamma / \pi}{\Gamma^2 + (\epsilon - \varepsilon_0 - m \omega)^2}, 
  \label{eq:Floquet_DOS}
\end{equation}
\noindent
which, despite the mathematical similarity, does \textit{not} lead to the result in the SCQME for the simple reason that, without the CR, the occupation of the QD trivially reads
\begin{equation}
  P(t) = \sum_m J_m^2\left(\frac{\de}{\hbar \omega}\right) = 1,
\end{equation}
\noindent
i.e., the electron is stuck in the QD and therefore, no current can be generated. 

From \eqr{eq:quasienergies}, however, we can immediately observe the similarity with the concept of quasi-energies, with the only key difference that CR introduces an imaginary part. In the semiclassical derivation, we have seen how the Fermi-Dirac distribution is evaluated at the instantaneous energy of the QD $\varepsilon(t)$. In the SCQME, we can consider \eqr{eq:digamma} as its natural extension to a quantum formalism, which is evaluated at the (complex) energies
\begin{equation}
  \mathcal{E}_m = \varepsilon_0 + m \hbar \omega - i h\Gamma.
\end{equation}
This is an obvious consequence of the CR, which makes the state in the QD metastable, and the eigenstates of the self-consistent time propagator in \eqr{eq:self_consistent_time_evol} are exponentially decaying.

From classical mechanics, moreover, we know that, at the steady state, an exponential decay can be modelled as a phase delay compared to driving the system. This phase lag can be derived directly from the master equation in \eqr{eq:pop_ME}, but it is also easily obtained by noticing that, for both the SME and SCQME, 
\begin{equation}
  \arg{Y_N} = \arctan{\frac{\Gamma}{N \omega}} = \frac{\pi}{2} - \phi_N
  \label{eq:phi_N}
\end{equation}
\noindent
is notably independent of $\varepsilon_0$. 
This last remark is particularly interesting because, \textit{electrically}, we can separate real and imaginary parts of the SEB admittance as
\begin{equation}
  Y = G_S + i \omega C_T,
\end{equation}
\noindent
where $C_T$ takes the name of the \textit{tunnelling} capacitance while $G_S = 1 / R_S$ is the Sisyphus conductance \cite{Vigneau_2023,Mizuta2017}. Equation~(\ref{eq:phi_N}) thus indicates that the distinction between the \textit{resistive} and \textit{reactive} response of the SEB is dictated uniquely by its dynamical properties (quasi-energies), and the detuning only modulates the \textit{magnitude} of the signal. In Appendix~\ref{sect_GC}, we discuss the significance of this on the energy balance of the system.

Because of the non-unitary interaction with the CR, it is formally impossible to write the SEB evolution as a \textit{single} ket. Section~\ref{General_ME} proves that the dynamics discussed thus far are the only solution to the SEB SCQME that satisfies the requirement of a density matrix, and therefore, is the steady state, no matter the initial electronic state \cite{Albert_Bradlyn_Fraas_Jiang_2016}. 

Nonetheless, we believe it valuable to the reader to stress that the metastable counterpart of the modes in \eqr{eq:Floquet_single} are the building blocks of the SEB evolution.
The interpretation of the dynamics in terms of damped-driven Floquet modes with equally spaced quasi-energies will lead to a novel interpretation of Power Broadening as an \textit{interference} phenomenon and will serve as an intuitive explanation of a novel effect introduced in this work: Floquet Broadening.

\begin{figure}
  \centering
  \includegraphics[width = 0.99 \linewidth ]{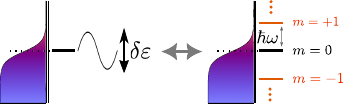}
  \caption{Pictorial representation of the dressed states in a driven QD. The single state in the QD whose energy is driven sinusoidally in time can be equivalently thought of as a \textit{ladder} of levels equally spaced by one photon energy, whose occupation is dictated by the strength of the drive.}
  \label{fig:fig_Floquet}
\end{figure}

\section{Reflectometry Lineshapes in Different Regimes}
\label{Regimes}

From \eqr{eq:Y_q}, it ought to be clear that four energy scales, namely $k_B T$, $\delta \varepsilon$, $\hbar \omega$, and $h \Gamma$, determine the SEB dynamics. In this section, therefore, we shall discuss the interplay between them and the different behaviour of the SEB when one dominates compared to the others. 
Finally, we will introduce the effect of photon loss $\kappa$ in the PhO and determine its repercussions on the effective admittance.
Considering the discussion of \eqr{eq:phi_N}, in this section, we shall only discuss the magnitude of the reflectometry signal $|Y|$. 
When possible, general properties of $Y_N$ are discussed, with a particular focus given to the fundamental frequency, $Y_1$, due to its experimental significance.

In particular, for the small-signal regime, it is possible to expand the Bessel functions as
\begin{equation}
  J_m(x) \xrightarrow[x\rightarrow0]{} \frac{x^m}{2^m m!}.
  \label{eq:J_exp_0}
\end{equation}
Therefore, when $\de \ll \hbar \omega$ we can neglect all $N>1$ terms. Physically, this is equivalent to the intuitive fact of considering only first dressed state in the Floquet ladder.

In this case, it is trivial to show that one always obtains
\begin{equation}
  Y_1 = \frac{(\alpha e)^2}{2 k_B T} \frac{\Gamma \omega}{\omega - i \Gamma} \mathcal{F}'(\varepsilon_0),
  \label{eq:Y1_small}
\end{equation}
\noindent
with $\mathcal{F}'$ a new function defined such that $\mathcal{F}'(\varepsilon_0) \in [0, 1]$ and, more notably, 
\begin{equation}
  \mathcal{F}'(\varepsilon_0 = 0) \xrightarrow[h \Gamma/k_B T \rightarrow 0]{\hbar \omega/k_B T \rightarrow 0} 1.
\end{equation}
Therefore, the different regimes only differ for the shape of $\mathcal{F}'(\varepsilon_0)$.
Interestingly, for a given temperature, the maximum achievable signal is $\frac{(\alpha e)^2}{2 k_B T}$, which we will see is the semiclassical prediction \cite{Vigneau_2023,oakes2022quantum}.
Perhaps unsurprisingly, we will also prove that, in all regimes, the maximum signal for the fundamental is always obtained at zero detuning.
It is interesting to point out that, as we shall see, $\mathcal{F}'$ is not necessarily an algebraic function of the excitation frequency $\omega$, and, thus, it is not generally possible to replace $Y_1(\omega)$ with a finite network of resistors, capacitors, and inductors, unlike in the semiclassical treatment \cite{Peri_2023}.

\subsection{Thermal Broadening}
\label{thermal_broadening}
The most straightforward regime is when $k_B T \gg \delta \varepsilon, \hbar \omega, h \Gamma$. In this case, the thermal smearing of the Fermi-Dirac distribution in the CR is the dominant effect. Therefore, all the assumptions of the SME are valid, and the SEB behaves fully semiclassically.

In particular, if $k_B T \gg \delta \varepsilon$, it is trivial to obtain, through Taylor expansion for small $\de$ of the SME in \eqr{eq:Y_sc}, the well-known result
\begin{equation}
  \begin{aligned}
  Y_1& = 2\mathcal{C}_1 \frac{\partial}{\partial \varepsilon_0 } f(\varepsilon_0) = \\
   &= \frac{(\alpha e)^2}{2 k_B T} \frac{\Gamma \omega}{\omega - i \Gamma} \cosh^{-2} {\parens{\frac{\varepsilon_0}{2 k_B T}}}
  \end{aligned}
\end{equation}
\noindent
as expected from \eqr{eq:Y1_small}.

From \eqr{eq:Y_q}, we can obtain the same result via \eqr{eq:J_exp_0}.
Therefore, in the limit $k_B T \gg h \Gamma$, $\mathcal{F}_{-m}^{-}(\varepsilon_0) \approx f(\varepsilon_0 - \hbar m \omega)$, and
\begin{equation}
  \begin{aligned}
  Y_1 = \frac{\mathcal{C}_1}{\de} \frac{\delta \varepsilon}{\hbar \omega} \parens{f(\varepsilon_0 + \hbar m \omega) - f(\varepsilon_0 - \hbar m \omega)}& \approx \\
   \approx  2 \mathcal{C}_1 \frac{\partial}{\partial \varepsilon_0 } f(\varepsilon_0), 
  \end{aligned}
  \label{eq:Y1_der}
\end{equation}
\noindent 
where the latter approximate equality derives from the fact that if $k_B T \gg \hbar \omega$ the difference approaches the definition of a derivative. 

In fact, in the small-signal regime, we can show by Taylor expansion of \eqr{eq:Y_sc} around $\delta \varepsilon = 0$ that \cite{oakes2022quantum}
\begin{equation}
  Y_N \propto \frac{\partial^N}{\partial \varepsilon_0^N } f(\varepsilon_0).
  \label{eq:YN_TB}
\end{equation}
We obtain a similar result for the SCQME, whereby recursively using \eqr{eq:Bessel_centerpoint}, $Y_N^{SCQME}$ follows the finite difference stencil coefficients for the $N$-th derivative with second-order accuracy. Thus, in the limit of $k_B T \gg \hbar \omega, \de$ the SCQME result approaches \eqr{eq:YN_TB}, as expected.

\subsection{Lifetime Broadening}
\label{Lifetime_Broad}

By remaining in the small-signal regime, it is interesting to compare the two models when $h \Gamma \gtrsim \kt \gg \de, \hbar \omega$. In this case, the QD is strongly coupled to the CR, and the reservoir-induced broadening of the (metastable) discrete level is comparable with thermal energy, know as Lifetime Broadening (LB).
This effect, however, is not considered in the SME. This should be apparent as $\Gamma$ and $\omega$ enter in \eqr{eq:Y_sc} only in the prefactor $\mathcal{C}_N$. Therefore, in the semiclassical picture, $\Gamma$ only \textit{rescales} the signal because of the high-pass filtering of the gate current from the series quantum capacitance,  but does not influence its \textit{lineshape}. 

\begin{figure}[h!]
  \centering
  \includegraphics[width = 0.99 \linewidth ]{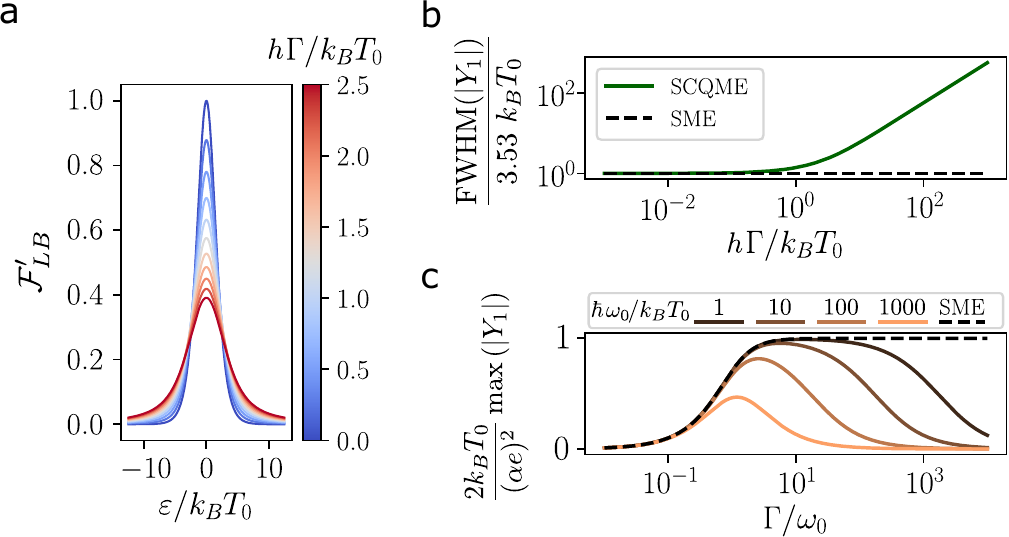}
  \caption{Effect of Lifetime Broadening (LB) on $Y_1$. (a) Shape of the function $\mathcal{F}'_{LB}(\varepsilon_0)$ (\eqr{eq:LB_convolution}) for varying $h \Gamma / k_B T_0$, showing how the broadening of the peak is accompanied by a lowering of the maximum. (b) FWHM of $|Y_1|$ showing the linear increase when $h \Gamma \gtrsim k_B T$. The graph is normalized to $3.53~k_B T$, the FWHM of $|Y_1|$ in the TB regime (SME). (c) Maximum of $|Y_1|$ normalized to the TB regime (SME). This panel shows the competition between the increase in signal because of $C_1$ and the drop because of the broadening of $\mathcal{F}'_{LB}$ for $h \Gamma \gg k_B T$.}
  \label{fig:fig2_LB}
\end{figure}

Using the mathematical result in Appendix~\ref{Digamma}, we can write the small-signal SEB admittance in the LB regime as \eqr{eq:Y1_small} with
\begin{equation}
  \mathcal{F}'_{LB}(\varepsilon_0) = \frac{2}{\pi^2} \Re\left[\psi_1\left(\frac{1}{2} + i \frac{\varepsilon_0}{2 \pi k_B T}+ \frac{h \Gamma}{2 \pi k_B T}\right)\right],
\end{equation}
\noindent
where $\psi_1(z)$ is the trigamma function. Notably,
\begin{equation}
  \begin{aligned}
  &\mathcal{F}'_{LB}(\varepsilon_0) = 
  \frac{\partial}{\partial \varepsilon} \mathcal{F}_0^{-}(\varepsilon_0) = \\
  &= \frac{1}{2k_B T}
  \parens{\cosh^{-2} {\parens{\frac{\epsilon}{2 k_B T}}} * \frac{\Gamma/\pi}{\Gamma^2 + (\epsilon - \varepsilon_0)^2 }}
\end{aligned}
\label{eq:LB_convolution}
\end{equation}
\noindent
which is the convolution between the $\cosh^{-2}$ in the TB regime to a Lorentzian peak in the case $h \Gamma \gg k_B T$, interpolating smoothly between the two. The effect of increasing tunnel rate in the admittance lineshape is shown in Fig.~\ref{fig:fig2_LB}a.

Interestingly, experimentally, it is a known fact that different transitions may show a different Full-Width-Half-Maximum (FWHM) at the same electron temperature. It has been shown that the small-signal quantum capacitance of a reservoir transition in the LB regime is equal to \cite{chawner2021nongalvanic,House_2015}
\begin{equation}
  \begin{aligned}
  C_Q = \frac{(e \alpha)^2}{2 k_B T} & \frac{1 }{1 + \omega^2/\Gamma^2}\cdot \\
  \cdot& \parens{\cosh^{-2} {\parens{\frac{\epsilon}{2 k_B T}}} * \frac{\Gamma/\pi}{\Gamma^2 + (\epsilon - \varepsilon_0)^2 }}.
  \end{aligned}
  \label{eq:cq_lb}
\end{equation}
\noindent
Therefore, considering the result in \eqr{eq:LB_convolution}, the self-consistent treatment in the SCQME naturally extends \eqr{eq:cq_lb} to both the Sisyphus resistance and higher harmonics. 

Fig.~\ref{fig:fig2_LB}b shows the former effect of LB on the Full-Width Half-Maximum (FWHM) of the peak, where we see that for $h \Gamma \gtrsim k_B T$, the FWHM starts increasing from the expected TB value of $3.53~k_B T$, obtained from the inverse hyperbolic cosine, and becomes linear with the tunnelling rate for increasing tunnel rates. In particular,
\begin{equation}
  \textnormal{FWHM}(|Y_1|) \xrightarrow[h \Gamma \gg k_B T, \hbar \omega, \de]{} 2 h \Gamma.
\end{equation}
For clarity, in all Figures, the parameters which are not varied are indicated with a subscript $0$.

Finally, we can consider the effect of $\Gamma$ on the amplitude of SEB admittance at the fundamental frequency, which is shown in Fig.~\ref{fig:fig2_LB}c. As discussed above, $\Gamma$ enters in $Y_1$ both in the prefactor and because of lifetime broadening. The former is a purely classical effect, and it is due to the charge requiring a finite time to tunnel out of the QD. In contrast, the other is purely quantum and arises from the effective density of states of the electron in the SEB. 
In Fig.~\ref{fig:fig2_LB}a, we show the effect purely of the latter for increasing $\Gamma/\omega$, which highlights another crucial effect of LB. 
As the coupling to the CR increases, the discrete energy level in the QD becomes increasingly broader, and thus the peak \textit{decreases}.
Therefore, any spectroscopic \textit{broadening} of the lineshape inevitably translates into \textit{lowering} the maximum signal, as it must be the case because of the properties of the convolution.

This effect, however, is complemented by the semiclassical contribution in $\mathcal{C}_1$. One could posit that larger coupling to the CR ought to translate in a higher probability of tunnelling events per rf cycle.
This effect is shown in Fig.~\ref{fig:fig2_LB}c, where the inclusion of the prefactor $\mathcal{C}_1$ leads \textit{monotonically increasing} signal for increasing $\Gamma$. 

However, a Self-Consistent quantum picture clearly shows that this is not the case. In the limit of $h \Gamma \gg k_B T$, the discrete level is so broadly distributed in energy that \textit{no} tunnelling events can be observed. 
Consequently, the maximum reflectometry signal begins to drop when $h \Gamma$ starts to become comparable with $k_B T$ and completely vanishes for $h\Gamma \gg \hbar \omega, k_B T$. Consistently with Fig.~\ref{fig:fig2_LB}a, this is accompanied by an increase of the FWHM of the peak, as evidenced by Fig.~\ref{fig:fig2_LB}b.

More specifically, while there is no simple analytical expression for the behaviour in Fig.~\ref{fig:fig2_LB}c, we can notice that $\mathcal{F}'_{LB} \propto \Gamma^{-1}$ for large $\Gamma$. Therefore, we similarly expect $\textnormal{max}\left(|Y_1|\right) \propto \Gamma^{-1}$ for $h\Gamma \gg \hbar \omega$.

\subsection{Floquet Broadening}
\label{Polaron_Broad}

The final small-signal regime presented in this work is a novel kind of broadening which arises from the rf photon energy being comparable with thermal and lifetime broadening. Similarly to lifetime broadening, this effect is purely quantum and cannot be captured semiclassically, as $\omega$ only appears in the already-discussed prefactor $C_N$. Interestingly, however, neither $\de$ nor $\omega$ appears separately in the SCQME, but only in the ratio $\frac{\de}{\hbar \omega}$. 
We can gain further insight by reconsidering the effective density of states in \eqr{eq:Floquet_DOS}, where the Floquet modes are equally spaced by the photon energy $\hbar \omega$. Therefore, the ratio $\frac{\de}{\hbar \omega}$ determines the \textit{highest} dressed state that will be reached by the semiclassical voltage swing. Moreover, from the discussion in Section~\ref{Floquet_Modes}, the reflectometry lineshape is determined by the interference between the different dressed states. 
Therefore, if $\hbar \omega \gtrsim h \Gamma$, i.e., the separation between the Floquet modes is comparable with the broadening of the levels and, thus, the discrete nature of the dressed states may emerge.

In this work, we present a novel result, which we name Floquet Broadening (FB), which becomes apparent in the regime where  $\hbar \omega \gtrsim \kt, h \Gamma$, i.e., the discrete nature of the photon energy becomes the primary source of \textit{broadening} in the SEB admittance. Here, we take the occasion to stress that FB is an \textit{inherently} quantum phenomenon, arising from the discrete spacing of the Floquet modes, thus, uncapturable with any semiclassical treatment.

\begin{figure}[htb!]
  \centering
  \includegraphics[width = 0.99 \linewidth ]{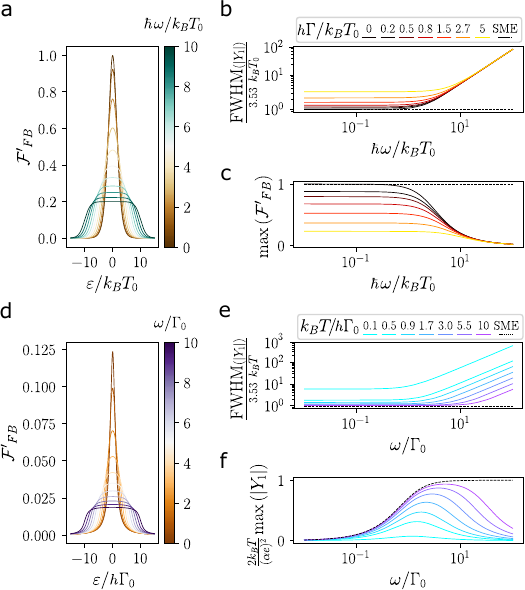}
  \caption{Effect of FB in for varying $\hbar \omega/k_B T_0$ (a-c) and $\omega/\Gamma_0$ (d-f). Panels (a,d) show how for increasing $\omega$, $\mathcal{F}'_{FB}$ approximates increasingly less a derivative and the consequent lowering of the peak accompanying the broadening.  Panels (b,e) show the increase of FWHM of $|Y_1|$, eventually reaching $2 \hbar \omega$, while panels (c,f) show the dependence of the maximum admittance with drive frequency.}
  \label{fig:fig4_DB}
\end{figure}

From a mathematical standpoint, FB stems from the fact that, for $\hbar \omega \sim k_B T$, \eqr{eq:Y1_der} becomes an increasingly worse approximation of the derivative. 
Therefore, we can use the small-signal expansion in \eqr{eq:J_exp_0} to write the admittance in the form of \eqr{eq:Y1_small} with 
\begin{equation}
  \begin{aligned}
  \mathcal{F'}_{FB} = \frac{k_B T}{\hbar \omega} \Im\bigg[&\psi_0\left(\frac{1}{2} + i \frac{\varepsilon_0 +\hbar \omega}{2 \pi k_B T}+ \frac{h \Gamma}{2 \pi k_B T}\right) -\\
  -&\psi_0\left(\frac{1}{2} + i \frac{\varepsilon_0 -\hbar \omega}{2 \pi k_B T}+ \frac{h \Gamma}{2 \pi k_B T}\right)\bigg].
\end{aligned}
\label{eq:F_FB}
\end{equation}
\noindent
Recalling the fact that $\psi_0$ is the digamma function, and $d/dz~\psi_0(z) = \psi_1(z)$, it is clear that for slow frequencies, $\mathcal{F'}_{FB} \approx \mathcal{F'}_{LB}$, while for larger photon energies, it is possible to resolve the two separate terms in \eqr{eq:F_FB}, which correspond to the first rotating and counter-rotating dressed states. 
We show the results in Fig.~\ref{fig:fig4_DB}, where we present the effect for the cases $ k_B T \gg h \Gamma$ (Fig.~\ref{fig:fig4_DB}a-c) and $h\Gamma \gg k_B T$ (Fig.~\ref{fig:fig4_DB}d-f).

In particular, it is interesting to consider the effect of increasing $\omega$ on the FWHM and maximum admittance in Fig.~\ref{fig:fig4_DB}b,e and Fig.~\ref{fig:fig4_DB}c,f, respectively.
Considering \eqr{eq:F_FB} and ignoring LB for simplicity, we expect because of the factor $k_B T / \hbar \omega$ in \eqr{eq:F_FB}
\begin{equation}
  |Y_1^{SCQME}| \propto \left|\mathcal{C}_1\right| \omega^{-1} \propto \frac{\Gamma}{ \sqrt{\Gamma^2 + \omega^2}}.
\end{equation}
In contrast, in the semiclassical case
\begin{equation}
  |Y_1^{SME}| \propto \left|\mathcal{C}_1\right| \propto \frac{\Gamma \omega}{ \sqrt{\Gamma^2 + \omega^2}}.
  \label{eq:omega_semicl}
\end{equation}
Therefore, contrary to the SME, the SCQME predicts a reduction in the SEB signal with increasing $\omega$ (Fig.~\ref{fig:fig4_DB}c,f).
Similarly to LB, this drop in signal is accompanied by an increase in the FWHM (Fig.~\ref{fig:fig4_DB}b,e). In particular, 
\begin{equation}
  \textnormal{FWHM}(|Y_1|) \xrightarrow[\hbar \omega \gg k_B T, h \Gamma, \de]{} 2 \hbar \omega.
\end{equation}

There is some interesting physical insight into the process of rf reflectometry to be gained from this regime. Semiclassically, one can picture increasing the measurement frequency in the small-signal regime as increasing the \textit{chances} per unit time of electron tunnelling. Therefore, this shall increase the maximum signal until $\omega$ becomes of the order of the tunnelling rate $\Gamma$. At that point, the probability of a tunnel event per unit cycle reads 
\begin{equation*}
P_{cycle} = 1 - e^{- \Gamma \tau_{cycle}} \approx \Gamma \tau_{cycle},
\end{equation*}
\noindent
where $\tau_{cycle} = 2 \pi / \omega$. Therefore, the average gate current reads
\begin{equation*}
  \braket{I_g} \propto e P_{cycle} /\tau_{cycle}
\end{equation*} 
\noindent
which, as shown in \eqr{eq:omega_semicl} and Fig.~\ref{fig:fig4_DB}f, saturates and becomes independent of frequency (dashed line).

From a quantum perspective, the tunnelling of the electron in (out) the CR is more akin to a \textit{measurement} of the state of the QD. Increasing $\omega$ means increasing the measurement rate, potentially much faster than the timescales of the (non-unitary) quantum dynamics. Therefore, the electron becomes \textit{trapped} inside (outside) the QD, a process resembling the Quantum Zeno effect \cite{gambetta_quantum_2008}.

Another interpretation may come from the Heisenberg uncertainty principle, as increasing $\omega$ would attempt to define the energy of a level at time scales faster than its lifetime. Therefore, we expect a \textit{Heisenberg} energy broadening of the order $2\pi/ \tau_{cycle} = \omega$, as shown in Fig.~\ref{fig:fig4_DB}.

Lastly, we note how FB is the only small-signal regime where, as anticipated, $\mathcal{F}'$ is \textit{not} an algebraic function of $\omega$. Therefore, it is not possible to express the SEB as equivalent to a \textit{finite} network of linear circuit component \cite{Foster_1924,Nigg_2012}, which is instead always the case with a semiclassical treatment of the radiation \cite{Peri_2023}. Therefore, this highlights the quantum nature of FB, which is only present in a theory that goes beyond the adiabatic (secular) limit.

A more detailed discussion of the optimal parameters for the SEB used as a sensor, taking into account LB and FB, is outlined in Appendix~\ref{Optimal_signal}.

\subsection{Power Broadening}
\label{Power_Broad}

As previously discussed, it is experimentally interesting to compare the two master equations in the large-signal regime, where $\delta \varepsilon \gtrsim k_B T, h \Gamma, \hbar \omega$. 
For reasons that shall become apparent, this is commonly referred to as the power-broadened (PB) regime, which we will begin to discuss in the case where $k_B T, h \Gamma \gg \hbar \omega$.

\begin{figure}[htb!]
  \centering
  \includegraphics[width = 0.99 \linewidth ]{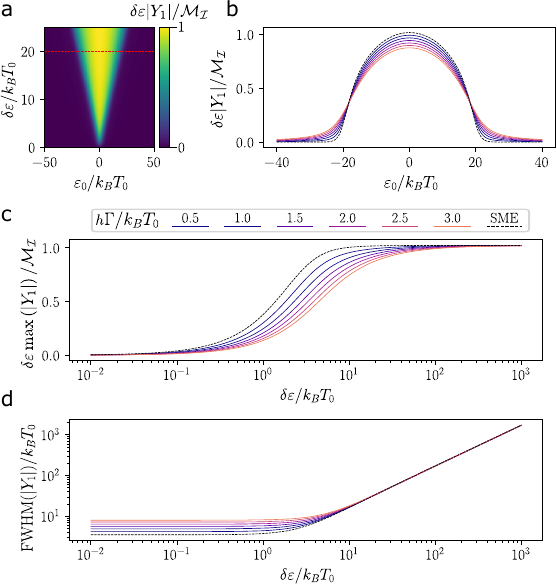}
  \caption{Effect of increasing the amplitude of the driving in the large-signal regime. (a) Map of the normalized gate current as a function of detuning offset and driving amplitude , showing the well-known PB fan. (b) Line cut of (a) (red dashes) showing the combined effect of PB and LB. (c) Saturation of the gate current in the limit of very large drive ($\delta \varepsilon \gg k_B T, h \Gamma, \hbar \omega$), and its retardation because of LB. (d) FWHM of $|Y_1|$, showing the convergence to $\sqrt{3} \delta \varepsilon$ in the limit of very large drive.}
  \label{fig:fig3_PB}
\end{figure}

In Fig.~\ref{fig:fig3_PB}a, we show how the SCQME can reproduce the familiar power-broadening fan predicted by the SME, where the admittance linewidth increases linearly with the amplitude of the detuning oscillations \cite{oakes_fast_PRX,oakes2022quantum,Vigneau_2023,cochrane2022intrinsic,Chatterjee_2018,Ivakhnenko_Shevchenko_Nori_2023,SHEVCHENKO2010}. From \eqr{eq:Y_q_1}, however, we can see how the SCQME does so by considering an \textit{infinite summation} over dressed states rather than an integral over a smoothly varying voltage.
We can gain more insight into the physical reasons behind this via the use of Floquet theory presented in Section \ref{Floquet_Modes}, where we showed how a single level oscillating with a large amplitude can be modelled as a ladder of equally spaced levels.

However, we have already pointed out how a simple application of \eqr{eq:Floquet_DOS}, obtained without the CR, does \textit{not} lead to the SCQME result in \eqr{eq:Y_q} (or \eqr{eq:Y_q_1}). 
The reasoning why is that the Floquet modes must combine in such a way to generate a \textit{real} occupation probability.
At any frequency different to $\omega$, two different Floquet modes will accumulate a different phase each rf cycle. When averaged out over multiple cycles, any interference becomes negligible and, therefore, one can separately describe the single dressed states, as in \eqr{eq:Floquet_DOS}. If the excitation frequency is the same as the measurement one (or its integer fraction), the different dressed states have a \textit{defined} phase relationship between each other, and will be able to interfere to generate the various harmonics of the driving.
The signal at the $N$-th harmonic is obtained by combining \textit{rotating} waves at $N \omega$ and \textit{counter-rotating} waves at $-N \omega$. 
Or, to phrase it differently, reflectometry techniques are sensitive to changes in AC currents (i.e., emitted and absorbed photons) and, thus, to \textit{transitions} between the ladder of dressed states. Therefore, this result can be considered an AC extension to the Tien-Gordon model \cite{Tien_Gordon_1963}, which is only concerned with direct (DC) conductance and, thus, time-averaged properties of the system \cite{Platero_Aguado_2004}.

All this is particularly apparent for the fundamental in \eqr{eq:Y_q_1}, where the lineshape is constructed from the \textit{interference} of the $m$-th and $-m$-th Floquet modes. This equation also highlights the well-known result in Floquet theory that \textit{all} the modes enter the definition of the system's response at every harmonic\cite{kohler_driven_2005,Rudner_Lindner_2020}.
From this discussion, it ought to be clear how PB is not merely due to incoherence with the CR but rather an \textit{interference} phenomenon between the Floquet modes of the damped-driven quantum system. The stochastic nature of the QD-CR interaction can thus be reinterpreted as a scattering process between the dressed states or as a finite lifetime of the polaron state.

Another interesting property of PB is how it is affected by LB (Fig. 5b-d). In the small-signal regime, LB arises as an additional broadening in addition to the thermal smearing of the reservoir. If $\de \gg k_B T \gg h \Gamma$, however, we can imagine that the voltage swing will dominate, and the QD will be emptied (filled) with probability 1 every cycle \cite{oakes2022quantum}. In the semiclassical picture, this will lead to a saturation of the gate current.
This trend is clearly shown in Fig.~\ref{fig:fig3_PB}c. 
Notably, the admittance, which is defined as the gate current over the excitation voltage, drops as $\de^{-1}$. Therefore, for  visual clarity, in Fig.\ref{fig:fig3_PB} we plot the product $\de |Y_1|$. 

Lifetime broadening, on the other hand, is an \textit{intrinsic} property of the QD and dictates the lifetime of \textit{all} dressed states. Therefore, we would expect it to play a role even in the regime where $\de \gg h \Gamma \gtrsim k_B T$. 
Indeed, in Fig.~\ref{fig:fig3_PB}b, we see that increasing $\Gamma$ leads to a broadening of the lineshape and the consequent lowering of the peak at constant power. However, by increasing the rf power, we can still reach saturation (Fig.~\ref{fig:fig3_PB}c). 
In the semiclassical picture, we can interpret this as a broadening of the density of states of the QD. Therefore, similarly to the previous argument regarding temperature, one can imagine the voltage swing to be large enough to ignore the intrinsic broadening of the level. If this is sufficient to fully empty or fill the QD every excitation cycle, we expect saturation to be reached. 
From a quantum perspective, we notice in \eqr{eq:kernel2} that the evolution of the polaron phase is proportional to $\de$, while it decays with $\Gamma$. 
Recalling the discussion in Section~\ref{Lifetime_Broad} (see also Appendix~\ref{sect_GC}), LB decreases the signal because the polaron dynamics is \textit{cut short}. Increasing the number of photons can accelerate the dynamics fast enough to overcome the finite lifetime. However, it comes at the expense of a broader peak because of Heisenberg (Fig.~\ref{fig:fig3_PB}d).

While a simple analytical solution is not available for the SME in \eqr{eq:Y_sc}, we can notice that in the regime $\de \gg k_B T \gg h \Gamma$ the integrand of the Fourier transform $f(\varepsilon(t))$ tends to a square wave. Thus, we can immediately write in the very large signal regime \cite{oakes2022quantum}
\begin{equation}
  \begin{cases}
  &Y_N^{SME} = \frac{2}{\pi}\frac{\mathcal{C}_N}{N \de} \sin{\left[ N~ \text{arccos}\left(\frac{\varepsilon_0}{\delta \varepsilon}\right)\right]} \hfill ~~|\varepsilon_0| < \de\\
  & Y_N^{SME} = 0 \hfill ~~|\varepsilon_0| \geq \de
\end{cases}
\label{eq:PB_SC}
\end{equation}
\noindent
which, for $N=1$, reduces to
\begin{equation}
  Y_1^{SME}\bigg|_{\de \rightarrow \infty} = \frac{2(\alpha e)^2}{\pi \de}\frac{\Gamma \omega}{\omega - i \Gamma} \sqrt{1 + \left(\frac{\varepsilon_0}{\delta \varepsilon}\right)^2 }.
  \label{eq:PB_SC_1}
\end{equation}

For ease of notation, we can therefore define 
\begin{equation}
  \mathcal{M_I} = \max\left(|I_1|\right)\bigg|_{\de \rightarrow \infty} = \frac{2(\alpha e)^2}{\pi} \left|\frac{\Gamma \omega}{\omega - i \Gamma}\right|,
\end{equation}
\noindent
which is the maximum gate current semiclassically achievable at the fundamental.
Notably, this saturates at very large power, and thus the admittance, which is defined as the gate current over the excitation voltage, drops as $\de^{-1}$. 

Interestingly, we shall note that \eqr{eq:PB_SC_1} (and \eqr{eq:PB_SC}) have a discontinuous derivative at $\varepsilon_0 = \pm \delta \varepsilon$. 
This can be understood by considering that, when thermal broadening is negligible, the Fermi-Dirac function tends to be a step-function with a discontinuous derivative. The smearing of the peak caused by LB, however, will smoothen these corners and keep every derivative continuous, as depicted in Fig.~\ref{fig:fig3_PB}b.

Finally, from Fig.~\ref{fig:fig3_PB}d, we notice an interesting trend. Because of the properties of \eqr{eq:Y_q_1}, the LB of the peak happens in such a way as to make FWHM only dependent on $\delta \varepsilon$. Therefore, we can use \eqr{eq:PB_SC_1} to derive the fact that, we have 
\begin{equation}
  \textnormal{FWHM}(|Y_1|) \xrightarrow[\de \gg k_B T, h \Gamma, \hbar \omega]{} \sqrt{3} \delta \varepsilon \approx 1.73 \delta \varepsilon
\end{equation}
\noindent
independently from $k_B T$ or $h \Gamma$.

\begin{figure}[htb!]
  \centering
  \includegraphics[width = 0.99 \linewidth ]{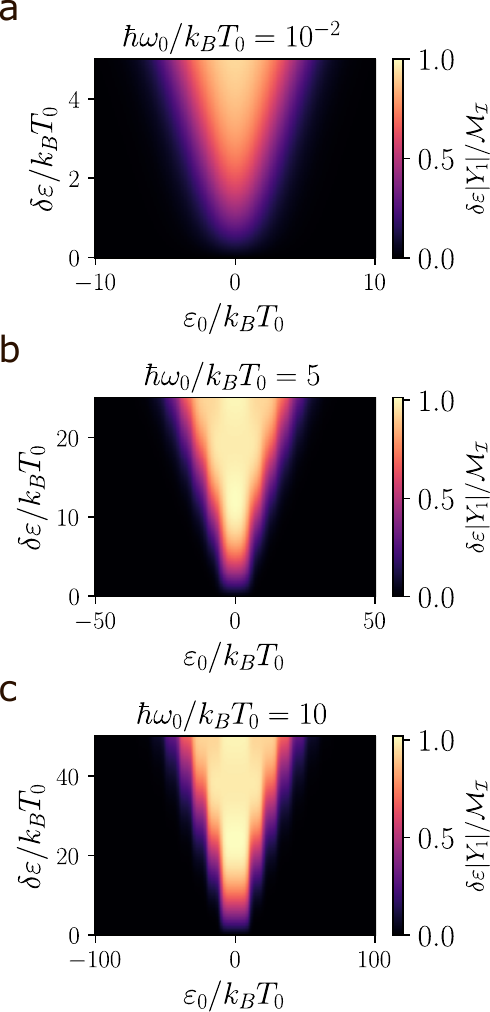}
  \caption{Evolution of the PB fan for increasing $\omega$. As we enter the FB regime, the effect of the \textit{discrete} summation in $Y_1$ becomes more clear, revealing the underlying ladder of dressed states. As this occurs, the gate current is permitted to go above the semiclassical limit, reaching $\de |Y_1| \approx 1.02 \mathcal{M_I}$ for $\de / \hbar \omega  \approx 2.40$.}
  \label{fig:fig_FLPB}
\end{figure}

Lastly, it is interesting to consider the effect of PB in conjunction with FB, in the case where $\de \gg \hbar \omega \gtrsim k_B T, h \Gamma$. Similarly to the small-signal regime, when the photon energy becomes significant with respect to the other broadenings, we expect to see the \textit{discrete} nature of the summation over the dressed states in \eqr{eq:Y_q_1}. We show the effect of increasing excitation frequency in Fig.~\ref{fig:fig_FLPB}, where it is clear how the PB fan becomes a more faithful depiction of the \textit{ladder} of dressed states. 

\textit{En passant}, it is interesting to note how, in the case of $\hbar \omega > k_B T, h \Gamma$ it is possible to achieve \textit{more} gate current that the semiclassical limit. It is possible to show, in fact, that for $\de / \hbar \omega  \approx 2.40$, $\de |Y_1| \approx 1.02 \mathcal{M_I}$. Although not a significant enhancement, it stands as testament of the coherent interference at the root of the process generating the reflectometry signals in this regime and points towards new experiments in the field of coherent hybrid quantum-classical circuits.

\subsection{Photon Loss Broadening}
\label{Measurement_broad}

Up until this point, we have considered the PhO as a perfect cavity held in a coherent state. However, since the SEB dynamics are purely dictated by the interaction between the ladder of dressed states, we must also consider photon loss, since this will cause the various dressed states to dephase and thus may strongly affect the signal. 
We can account for this effect by assuming the loss rate $\kappa$ to be small compared to the rf frequency $\omega$. Therefore, redefining the PhO Hamiltonian as

\begin{equation}
  H_{PhO} = \hbar \omega a^\dagger a + \hbar \kappa \mathcal{D}[a],
\end{equation}
\noindent
which, in conjunction with a classical drive supplied by the rf source, will cause the photon number to fluctuate in time as
\begin{equation}
  n(t) = \bar{n} + \delta n(t).
\end{equation}

For a damped-driven system, the average number of photons reads \cite{gambetta_qubit-photon_2006}
\begin{equation}
  \bar{n} = \frac{\de^2}{(2 g)^2 + \hbar^2 \kappa^2},
\end{equation}
\noindent
while the time correlation of the fluctuations obeys \cite{PhysRevA.69.062320}
\begin{equation}
  \langle \delta n(t) \delta n(t') \rangle = \bar{n} e^{-\frac{\kappa}{2} |t-t'|} .
\end{equation}

With this additional term, we can write the Displacement Operator as \cite{doi:10.1063/5.0091953,Dutra_1997,Roccati_2022}
\begin{equation}
  D_{\alpha} (t) =  e^{\frac{|\alpha|^2}{2} (1 - e^{-\frac{\kappa}{2} t})}  e^{-\frac{1}{2} \kappa a^\dagger a t}  D_{e^{i \omega t} \alpha}.
\end{equation}

In the Gaussian approximation for the phase, we can modify the kernel in  \eqr{eq:kernel2} as \cite{Dutra_1997,gambetta_quantum_2008,gambetta_qubit-photon_2006}
\begin{equation}
  \begin{aligned}
  &\mathcal{K}( \epsilon, t) =
e^{ i\left(\frac{\delta \varepsilon }{\hbar \omega}\right) \sin{\omega t}}
\sum_{m= - \infty}^{+ \infty} 
J_m \left( \frac{\delta \varepsilon}{\hbar \omega}\right)
e^{ -i m\omega t} \cdot \\&
\cdot\int_0^{+ \infty} d\tau 
e^{- i(\epsilon - \varepsilon_0 - m \omega) \tau} 
e^{- \Gamma \tau} 
e^{\bar{n}\left(1-\frac{\kappa}{2} \tau - e^{-\frac{\kappa}{2} \tau}\right)}.
\end{aligned}
\label{eq:kernel_mib}
\end{equation}

If we assume that $\kappa \gg \Gamma$, the lifetime of an electron in the QD is much longer than the coherence of the radiation. In this case, we can neglect the term $e^{-\frac{\kappa}{2} \tau}$, and it ought to be apparent how the \textit{effective} lifetime of the ladder of dressed states reads
\begin{equation}
  \tilde{\Gamma} = \Gamma + \kappa \frac{\de^2}{(2 g)^2 + \hbar^2 \kappa^2} = \Gamma + \gamma \left(\frac{\de}{\hbar \omega}\right)^2,
  \label{eq:gamma_loss}
\end{equation}
\noindent
which, for large input powers, is dominated by photon dynamics rather than electronic ones. As seen in the previous sections, it is helpful to describe the dynamics in terms of the ratio $\de/\hbar \omega$. Thus, in \eqr{eq:gamma_loss}, we have defined the rate
\begin{equation}
  \gamma = \frac{\kappa \omega^2}{(2 g / \hbar)^2 + \kappa^2}
  \label{eq:gamma_PLB}
\end{equation}
\noindent
to quantify the effect of photon loss to the CR in the SEB lifetime.

If, on the other hand, $\kappa \ll \Gamma$, photon loss becomes a small perturbation, and 
\begin{equation}
e^{\bar{n}\left(1-\frac{\kappa}{2} \tau - e^{-\frac{\kappa}{2} \tau}\right)} \approx e^{- \frac{\bar{n}\kappa^2}{8} \tau^2},
\end{equation}
\noindent
which means that the \textit{spectrum} of the SEB becomes the convolution of a Lorentzian and a Gaussian (i.e., a Voigt profile), the inhomogeneous part directly arising from the Poisson statistics of photon counts. Thus, this phenomenon has also been described as photon shot noise in the context of superconducting flux qubits \cite{Yan_Gustavsson_2016,Blais_2004,Sears_2012}.

We will refer to this novel effect as Photon Loss Broadening (PLB), and it is the SEB equivalent of the well-known phenomenon in circuit QED of measurement-induced dephasing \cite{gambetta_quantum_2008,gambetta_qubit-photon_2006,malekakhlagh_time-dependent_2022,Yan_Gustavsson_2016,slichter_measurement_induced_2012}. Once again, this result stresses how PB in a SEB is a \textit{coherent} process between dressed states in the QD.

We can analytically evaluate the integral in \eqr{eq:kernel_mib} and expand the resulting spectrum in an infinite series of Lorentzians \cite{gambetta_qubit-photon_2006}. Therefore, we can use the result in Appendix~\ref{Digamma} to introduce the effect of PLB simply by redefining 

\begin{equation}
  \mathcal{F}_m^{\pm}(\varepsilon_0)= \frac{1}{2} +\frac{e^{\bar{n}}}{2\pi} \sum_l  \frac{(-\bar{n})^l}{l!}\Im\left[\psi_0\left(\frac{1}{2} + \tilde{\Theta}^{\pm}_{l,m} \right)\right],
\end{equation}
\noindent
where
\begin{equation}
  \Theta^{\pm}_{l,m} = h\frac{\tilde{\Gamma} + l \kappa}{2 \pi \kt} \pm i\frac{\varepsilon_0 + m \hbar \omega}{2 \pi \kt}.
\end{equation}

In usual experimental settings, we have that $\hbar \kappa \ll g \ll \hbar \omega \sim h \Gamma$.
Therefore, we could approximate 

\begin{equation}
  \tilde{\Theta}^{\pm}_{l,m} \approx \tilde{\Theta}^{\pm}_{m} = \frac{h\tilde{\Gamma}}{2 \pi \kt} \pm i\frac{\varepsilon_0 + m \omega}{2 \pi \kt},
\end{equation}
\noindent
which allows us to sum the infinite series as 
\begin{equation}
  \mathcal{F}_m^{\pm}(\varepsilon_0) \approx \frac{1}{2} +\frac{1}{2\pi} \Im\left[\psi_0\left(\frac{1}{2} + \tilde{\Theta}^{\pm}_{m} \right)\right],
\end{equation}
\noindent
which shows how the main effect of PLB is a super-linear LB and the consequent reduction in the SEB signal.

\begin{figure}[htb!]
  \centering
  \includegraphics[width = 0.99 \linewidth ]{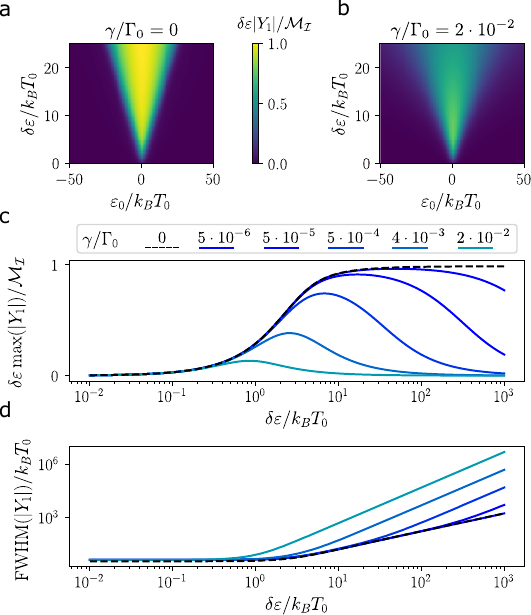}
  \caption{Effect of PLB. (a,b) Gate current maps showing the alteration of the PB fans in the presence of photon loss. Both maps are normalized to the maximum semiclassical gate current $\mathcal{M_I}$, showing how in the case of PLB this is never reached, and the signal starts decreasing for very large $\de$. (c,d) Maximum normalized gate current and FWHM of the peak for increasing $\gamma$. The effect of PLB causes panel (c) to not be monotonically increasing, causing a reduction of signal for increasing power. This is accompanied by a \textit{superlinear} increase in FWHM, which eventually become proportional to $\gamma \de^2$ (panel (d)).}
  \label{fig:fig6_PLB}
\end{figure}

We capture this effect in Fig.~\ref{fig:fig6_PLB}, which shows a comparison of the PB fan with $\gamma =0$ (Fig.~\ref{fig:fig6_PLB}a) and with $\gamma =0.2\cdot 10^{-2} \Gamma$ (Fig.~\ref{fig:fig6_PLB}b). Perhaps counterintuitively, in the case of a lossy cavity (resonator), increasing the input power does not always lead to an increase in the reflectometry signal, as observed experimentally \cite{oakes_fast_PRX,oakes2022quantum,vonhorstig2023multimodule}. However, we shall note that in an experimental setting, the degradation of the signal because of PLB may be complemented by heating the sample and/or the resonator, which may also lower the total signal. 
We highlight the effect of PLB in Figs.~\ref{fig:fig6_PLB}c-d, where we show maximum signal and its FWHM are shown as a function of detuning amplitude for varying $\gamma$. For small $\gamma$, the effect of PLB only manifests at very large detuning amplitudes, where we see a super-linear broadening for increasing $\de$ and the consequent reduction in the peak height. Roughly, we can estimate the threshold for this change in regime as 
\begin{equation}
  \de \gtrsim \hbar \omega \sqrt{\frac{\Gamma}{\gamma}},
  \label{eq:PLB_thr}
\end{equation}
\noindent
where PLB becomes comparable with the usual LB (which is only caused by the CR).
As shown in Section~\ref{Power_Broad}, the onset of PB is for 
\begin{equation}
\de \gtrsim k_B T, h \Gamma.
\label{eq:PB_thr}
\end{equation}
Therefore, depending on the value of $\kappa$, \eqr{eq:PLB_thr} may be reached before \eqr{eq:PB_thr}. Observing Fig.~\ref{fig:fig6_PLB}c, in this case, the maximum theoretical signal is never reached. Finally, we see that the FWHM of the peak increases as $\de^2$ (Fig.~\ref{fig:fig6_PLB}d). 

\section{Conclusions and Outlook}

We have presented a new theory to go beyond the semiclassical models for the development of equivalent circuits of quantum systems and applied it to a simple yet technologically useful QD system: the SEB. 
Our results reproduce the known behaviour of the SEB at low frequency and high temperature \cite{Cochrane2022,cochrane2022intrinsic}, and present an extension beyond the adiabatic regime. The self-consistent nature of our approach also allows us to include the effect of the finite lifetime of an electron in a QD when coupled to a reservoir, including the description of its resistive component, which was missing from the literature \cite{chawner2021nongalvanic,House_2015,Ahmed2018}. 

The mathematical formalism employed in this work sheds new light on the concept of gate current and the process of rf reflectometry, which is interpreted as the dynamics of polaron states in the QD. This allows us to see the well-known phenomenon of Power Broadening from a new angle, as an interference effect between dressed states in the QD.

Our theory extends the adiabatic semiclassical SEB description into a novel regime, which we call Floquet Broadening, where the main source of energy uncertainty comes from the discrete nature of the photonic part of the polaron. 
In doing so, our theory recovers the Heisenberg uncertainty principle, which is not respected by the semiclassical theory. 
In its discussion, we propose measurements which are technologically within reach for the experimental validation of our theoretical results, including a modification of the admittance lineshape and a drop of its maximum value at high drive frequencies.

The self-consistent quantum formalism here developed is fully general and applicable to quantum systems such as multi quantum dots or multi-level systems such as spin qubits. Therefore, it lays the foundations for developing fully quantum circuit equivalent models for quantum systems. The possibility of properly accounting for relaxation and dephasing phenomena beyond the adiabatic limit comes with the enticing perspective of extending such models to systems conceived for quantum computation, allowing for a more informed engineering of their design and operation.

Moreover, our quantum description of the photon source allows for the theoretical treatment of its non-idealities. This possibility has led us to introduce a new large-signal phenomenon, photon loss broadening, which can contribute to the experientially observed drop in SEB signal at very large powers. 
This is the first description in a QD system and terms of an equivalent circuit model of a well-known phenomenon observed in superconducting devices, known as photon shot noise or measurement induced dephasing \cite{gambetta_qubit-photon_2006,gambetta_qubit-photon_2006,malekakhlagh_time-dependent_2022,Yan_Gustavsson_2016,Blais_2004,Sears_2012,slichter_measurement_induced_2012}. 
Also in this case we present a clear experimental path for the measurement of this effect.

Finally, we have shown how, by exploiting quantum coherence in the system, the semiclassical limits of the maximum achievable AC currents can be overcome. While our example only presents a minor improvement (2\%), leveraging quantum phenomena to augment well-known devices may put within technological reach semiclassically unachievable goals.

\section*{Acknowledgements}

The authors acknowledge F.~Martins from Hitachi Cambridge Laboratory for useful discussions.
This research was supported by European Union's Horizon 2020 research and innovation programme under grant agreement no.\ 951852 (QLSI), and by the UK's Engineering and Physical Sciences Research Council (EPSRC) via the Cambridge NanoDTC (EP/L015978/1).
L.P. acknowledges support from The Winton Programme for the Physics of Sustainability for funding.
M.F.G.Z. acknowledges a UKRI Future Leaders Fellowship [MR/V023284/1]. 
L.C. acknowledges support from EPSRC Cambridge UP-CDT
(EP/L016567/1).

\appendix

\section{Master-Equation Derivation}
\label{app:General_ME}

To derive \eqr{eq:pop_ME}, it is fruitful to examine the problem in the Fock-Liouville space by considering the flattened density matrix \cite{manzano_short_2020}
\begin{equation}
  \rho \rightarrow \vec{\rho}_{FL} = \begin{pmatrix}
    P ,&
    C,&
    C^*, &
    1-P
  \end{pmatrix}^T.
\end{equation}

This picture allows us to explicitly write the LME in \eqr{eq:LME_main} in its matrix form, which, after some algebra~\cite{amshallem2015approaches} and making use of \eqr{eq:Fermionc_Character}, reads
\begin{widetext}
\begin{equation}
  \frac{d}{dt} \begin{pmatrix}
    P \\
    C\\
    C^* 
    \\ 1-P
  \end{pmatrix} = \begin{pmatrix}
    -\Gamma_+(t) & 0& 0 & \Gamma_-(t)\\
    0 & - \Gamma/2 - i \varepsilon(t)/\hbar & 0&0\\
    0  & 0& - \Gamma/2 + i \varepsilon(t)/\hbar & 0\\
    \Gamma_+(t) & 0& 0 & -\Gamma_-(t)
  \end{pmatrix}
  \begin{pmatrix}
    P \\
    C\\
    C^* \\ 
    1-P
  \end{pmatrix}.
  \label{eq:LME_mat}
\end{equation}
\end{widetext}

After matrix multiplication, we arrive at an expression for the coherence 
\begin{equation}
  \frac{d}{dt} C = \left(- \frac{\Gamma}{2} - i \frac{\varepsilon(t)}{\hbar}\right) C. 
  \label{eq:c_ME}
\end{equation}

Intuitively, \eqr{eq:c_ME} shows how the CR exponentially destroys any possible coherence between the full and empty QD. The system becomes a \textit{statistical} mixture of the two kets because of the \textit{stochastic} tunnelling events to and from the CR. Finally, we arrive at the key result that, generally, the charge dynamics are given by the statistical properties of $P(t)$, which are uniquely determined by the functional form of $\Gamma_-(t)$,

\begin{equation} 
  \frac{d}{dt} P + \Gamma P = \Gamma_- (t)
  \label{eq:app_pop_ME}
\end{equation}
\noindent
which is the result presented in Section~\ref{General_ME}.

\section{Derivation of the Tunnel Rates}
\label{app:Tunnel_Rates}

In the previous section, we showed how the tunnelling rate $\Gamma_-(t)$ fully defines the charge dynamics. In this section, we derive such tunnel rates in the semiclassical and self-consistent quantum formalism. In particular, we highlight the non-trivial approximations made in the semiclassical model and stress the comparison with the fully quantum formalism.

\subsection{Semiclassical Lindblad Master Equation}
\label{app:Semiclassical_gamma}

In this section, however, we shall briefly retrace the steps to derive a LME from \eqr{eq:H_sc2} to highlight the differences between the semiclassical result and the new self-consistent master equation derived in section \ref{self_ME}.

By defining the Liouville superoperators as
\begin{align}
  &\mathcal{L} (t) = \left[ H_0(t)+ H_R (t), \cdot \right]\\
  &\mathcal{L}'(t) = \left[ H_{DR} (t) , \cdot \right],
\end{align}
\noindent
we can write the semiclassical LME in the known form of \eqr{eq:ME_gen} where \cite{Yan_1998,Sowa_Mol_Briggs_Gauger_2018}
\begin{equation}
  \mathcal{G}(t - \tau) = \mathcal{T} e^{-i \int_t^\tau \mathcal{L} (t') dt' }
\end{equation}
\noindent
is the free propagator of the unperturbed dynamics, and $\braket{\cdot}_R$ represents the partial trace over the CR. The notation $\mathcal{T}$ represents the time-ordered integral.

We shall note that in \eqr{eq:ME_gen}, we have already assumed Markovian dynamics when extending the upper bound of the integral to $\tau \rightarrow + \infty$. As usual in the Lindblad formalism, we can compute the partial trace in \eqr{eq:ME_gen} via the Born approximation by requiring that the dynamics in the reservoir be faster than the CR-QD interaction timescales, and thus the CR can be considered constantly in thermal equilibrium. 
Therefore, 
\begin{align}
  & \braket{c_\epsilon}_R = \braket{c_\epsilon^\dagger}_R = 0\\
  & \braket{c_\epsilon^\dagger c_\epsilon}_R = f(\epsilon) \\
  & \braket{c_\epsilon c_\epsilon^\dagger}_R = f(-\epsilon) = 1-f(\epsilon),
\end{align}
\noindent
where 
\begin{eqnarray}
  f(\epsilon) = \frac{1}{e^{\epsilon/k_B T} + 1}
\end{eqnarray}
is the Fermi-Dirac distribution. After making what is commonly referred to as the Rotating Wave Approximation \cite{kohler_driven_2005,cochrane2022intrinsic,Sowa_Mol_Briggs_Gauger_2018,Sowa_Lambert_Seideman_Gauger_2020,Esterli2019,Mizuta2017,Vigneau_2023,kohler_floquet-markovian_1997,kohler_dispersive_2018} or the Instantaneous Eigenvalues Approximation \cite{Yamaguchi_Yuge_Ogawa_2017, Childs_Farhi_Preskill_2001}, \eqr{eq:ME_gen} becomes equivalent to \eqr{eq:LME_main}, with 

\begin{eqnarray}
  \Gamma_{\pm} (t) = \Re\left[\sum_\epsilon |V_\epsilon|^2 f(\pm \epsilon) \int_0^{\infty} d \tau e^{- i(\epsilon + \varepsilon(t) ) \tau}\right].
  \label{eq:gamma_pm_sc}
\end{eqnarray}

We can now take the continuous limit of the CR ($\sum_\epsilon \rightarrow \int \mathcal{D}(\epsilon) d\epsilon$, with $\mathcal{D}(\epsilon)$ the density of states in the CR) and take the wide-band limit of both $\mathcal{D}(\epsilon)$ and $V_\epsilon$ being weakly dependent on $\epsilon$. In this limit, \eqr{eq:gamma_pm_sc} becomes the well-known result in \eqr{eq:sc_rates}.
Notably, this is effectively equivalent to a stationary phase approximation of \eqr{eq:gamma_pm_sc}, and would be exact if $\varepsilon$ does not depend on time. In the driven case, the instantaneous eigenvalues approximation implicitly assumes an \textit{adiabatic} (secular) approximation of the propagator, and thus the evolution of the quantum phase, with $\omega$ much slower than any other timescale (i.e., $\Gamma$) \cite{Brasil_Fanchini_Napolitano_2013,Dann_Levy_Kosloff_2018,Yamaguchi_Yuge_Ogawa_2017,kohler_floquet-markovian_1997}. This will become apparent in the further discussion of the effective admittance.
Not only it immediately proves the ansatz in \eqr{eq:Fermionc_Character}, but it also defines $\Gamma = |V|^2\mathcal{D}$ \cite{Sowa_Mol_Briggs_Gauger_2018,Sowa_Lambert_Seideman_Gauger_2020}. 
It ought to be apparent now how we should expect \eqr{eq:Fermionc_Character} from the fermionic character of the CR and thus the identity $\{c,c^\dagger\} =1$. 

\subsection{Self-Consistent Single Electron Box Master Equation}
\label{app:self_ME}

In this section, we derive a fully quantum master equation, now treating the PhO quantum mechanically. To do so, we begin by re-defining the Liouvile superoperators as 

\begin{align}
  &\mathcal{L} (t) = \left[ H_{QD}+ H_R + H_{PhO} , \cdot \right]\\
  &\mathcal{L}'(t) = \left[ H_{DR} + H_{DP} , \cdot \right].
\end{align}

Therefore, the LME is now written in a similar form to \eqr{eq:ME_gen}, as 

\begin{equation}
\begin{aligned}
  \hbar \frac{d}{dt} \rho& =  -i \left[ H_{QD}(t), \rho(t) \right] - \\
  &-\int_0^{+ \infty} d\tau \langle \mathcal{L}'(t) \mathcal{G}(t, \tau) \mathcal{L}'(\tau)\mathcal{G}^\dagger(t, \tau) \rangle \rho(t),
\end{aligned}
  \label{eq:ME_q}
\end{equation}
\noindent
where the partial trace must now be taken over both the CR and the PhO. 

\subsubsection{Self-Consistent Propagator}
\label{app:self_cons_prop}

In Section \ref*{Semiclassical_gamma} and Appendix~\ref{app:Semiclassical_gamma}, we saw how the SME does not consider level broadening.  The reason for this is the combination of the instantaneous eigenvalues approximation with \eqr{eq:ME_gen} using the \textit{free} propagator $\mathcal{G}(t)$ in the Born approximation. We can remedy this by taking into account the self-consistent Born approximation, where we replace $\mathcal{G}(t)$ with a \textit{self-consistent} propagator \cite{Lee_2009}

\begin{equation}
  \mathcal{U}(t, \tau) = \mathcal{T} e^{-i \int_t^\tau (\mathcal{L} +\mathcal{L}')  (t') dt' }.
\end{equation}

We now consider the operators in the Heisenberg Picture \cite{Esposito_Galperin_2010,Hou_Wang_Wang_Ye_Xu_Zheng_Yan_2015} 

\begin{equation}
  \mathcal{U}(0, t)[d] = \braket{e^{-i \tilde{H} t} d  e^{-i \tilde{H} t}},
\end{equation}
\noindent
with $\tilde{H}$ representing the \textit{full} polaron-transformed Hamiltonian and the partial trace taken over the CR and PhO \textit{after} the time evolution. Using the BCH theorem, one can prove that this is equivalent to requiring \cite{Sowa_Mol_Briggs_Gauger_2018,Sowa_Lambert_Seideman_Gauger_2020}
\begin{equation}
  \begin{cases}
  & \hbar \frac{d}{dt} d(t) = i \tilde{\varepsilon}_0 d(t) -i \sum_{\epsilon} V_\epsilon D^\dagger(t) c_\epsilon(t)\\
  & \hbar \frac{d}{dt} c_\epsilon(t) = i \epsilon  c_\epsilon(t)- i V^*_\epsilon D(t) d(t) 
\end{cases}
  \label{eq:time_evol}
\end{equation}
\noindent
and that $D(t)$ evolves normally, as the phonon operators only appear in the free Hamiltonian $H_{PhO}$. Interestingly,  the time evolution of the ladder operators of the QD and CR are now coupled, reflecting how, for finite $V_\epsilon$, the state in the QD is metastable. Taking the Laplace transform of \eqr{eq:time_evol} in the wide-band limit of the CR yields the intuitive result \cite{Sowa_Lambert_Seideman_Gauger_2020,Sowa_Mol_Briggs_Gauger_2018}
\begin{align}
  &\mathcal{U}(0, t)[d] = d e^{-i \tilde{\varepsilon}_0 t} e^{- \Gamma t} \\
  &\mathcal{U}(0, t)[d^\dagger] = d^\dagger e^{i \tilde{\varepsilon}_0 t} e^{- \Gamma t}.
\end{align}

\subsubsection{Self-Consistent Tunnel Rates}
\label{app:self_cons_tunnel_rates}

Using the result in \eqr{eq:self_consistent_time_evol}, we can write the LME in the polaron frame as 
\begin{equation}
\begin{aligned}
 \hbar &\frac{d}{dt} \tilde{\rho}  = -i \left[ \tilde{H_S}, \tilde{\rho}(t) \right] - \\ 
 & -\int_0^{+ \infty} d\tau \left\langle 
\tilde{\mathcal{L}}'(t) \mathcal{U}(t, \tau) \tilde{\mathcal{L}}'(\tau)\mathcal{U}^\dagger(t, \tau) \right \rangle \tilde{\rho}(t),
\end{aligned}
\end{equation}
\noindent
which we can similarly recast in the form of \eqr{eq:LME_main} with the equivalent of \eqr{eq:gamma_pm_sc} reading
\begin{align}
  & \Gamma_{-} (t) = \frac{\Gamma}{\pi} \Re\left[\int d \epsilon f(\epsilon) \mathcal{K}(\epsilon, t) \right] \\
  & \Gamma_{+} (t) = \frac{\Gamma}{\pi}  \Re\left[\int d \epsilon (1-f(\epsilon)) \mathcal{K}^*(\epsilon, t)\right],
  \label{eq:gamma_pm_quantum}
\end{align}
\noindent
where we have defined 
\begin{equation}
  \mathcal{K}(\epsilon, t) = \int_0^{\infty} d \tau e^{- i(\epsilon -\varepsilon_0 ) \tau - \Gamma \tau} \braket{D(t)D^\dagger(\tau-t)}_{PhO}.
  \label{eq:kernel}
\end{equation}

To compute the partial trace over the PhO, we notice that the semiclassical result in \eqr{eq:field} can be obtained simply assuming that the radiation is in a coherent state $\ket{\de/2g}$. This view, although simplistic, can be justified by considering the typical phase noise characteristic of rf sources, together with the fact that, as made evident in, \eqr{eq:kernel}, in the self-consistent picture, phase-coherence is only required on timescales of the order $1/\Gamma$ ~\footnote{Typical figures for phase noise for a 1GHz Local Oscillator are of -100\,dBc at 100\,Hz offset and -40\,dBc at 1\,Hz offset, while experimentally $\Gamma$ varies between 10\,MHz and 100\,GHz. Thus, the excitation can be considered monochromatic for experimental purposes.}. We take the occasion also to stress that the Markovian assumption becomes natural in the self-consistent Born approximation, thanks to the self-consistent propagator decaying exponentially on the same timescales. 

In this case, we can use the well-known composition properties of displacement operators to write
\begin{equation}
  \begin{aligned}
  &\braket{D(t)D^\dagger(\tau-t)}_{PhO} = 
  \exp\parens{-\parens{\frac{g}{\hbar \omega}}^2(1-e^{-i \omega \tau})}\cdot\\
  &\cdot\exp\parens{-i\left(\frac{\delta \varepsilon }{\hbar \omega}\right) \parens{\sin{\omega (\tau - t)} +\sin{\omega t}}},
\end{aligned}
\end{equation}
\noindent
where we have separated the contribution of photon absorption and stimulated emission, containing both rotating and counter-rotating terms, and spontaneous emission, containing only counter-rotating terms independent of the microwave amplitude. We shall note that the spontaneous emission tends to 1 in the regime $g \ll \hbar \omega$. This is the case of weak QD-PhO coupling, which is the case of all the results we shall consider in this work. Therefore, for the sake of clarity and ease of notation, we shall disregard this term from now on.
We can compute the integral in \eqr{eq:kernel} by making use of the Jacobi-Anger identity
\begin{equation}
  \exp\left( -i x\sin{\omega \tau} \right) = \sum_{m= - \infty}^{+ \infty} J_m \left( x\right) e^{-i m \omega \tau},
\end{equation}
\noindent
where $J_m \left( x\right)$ is the Bessel function of the first kind. This solution allows us to write, neglecting spontaneous emission,
\begin{equation}
  \begin{aligned}
  \mathcal{K}( \epsilon, t) =
e^{ i\left(\frac{\delta \varepsilon }{\hbar \omega}\right) \sin{\omega t}}
&\sum_{m= - \infty}^{+ \infty} 
J_m \left( \frac{\delta \varepsilon}{\hbar \omega}\right)
e^{ -i m\omega t} \cdot \\&
\cdot\int_0^{+ \infty} d\tau 
e^{- i(\epsilon - \varepsilon_0 - m \omega) \tau} 
e^{- \Gamma \tau}.
\end{aligned}
\label{eq:kernel2}
\end{equation}
Notably, modulo the self-consistent term $e^{- \Gamma \tau}$, this expression resembles previous results obtained in the non-adiabatic quantum master equation framework or Floquet-Lindblad formalism \cite{Dann_Levy_Kosloff_2018,Yamaguchi_Yuge_Ogawa_2017,kohler_floquet-markovian_1997,Ikeda_Chinzei_Sato_2021,Mori_2023,Benito_Mi_Taylor_Petta_Burkard_2017}.
However, our treatment not only allows us to include the effect of the finite QD lifetime in the dynamics but immediately gives us a physical interpretation of the coefficients, as determined from the photonic part of the polaron, rather than having to derive them separately from a Floquet decomposition \cite{Dann_Levy_Kosloff_2018,Ikeda_Chinzei_Sato_2021,Mori_2023,Koski_2018}.

Exchanging the order of integration, we can write
\begin{equation}
  \begin{aligned}
  \Gamma_\pm(t) = \Gamma
\Re\bigg[
e^{ i\left(\frac{\delta \varepsilon }{\hbar \omega}\right) \sin{\omega t}}
\sum_{m= - \infty}^{+ \infty} &
J_m \left( \frac{\delta \varepsilon}{\hbar \omega}\right) \cdot\\
&\cdot e^{ -i m\omega t}
\mathcal{F}_m^{\pm}(\varepsilon_0)
\bigg],
  \end{aligned}
\label{eq:Gamma_pm_scq_long}
\end{equation}
\noindent
where we have defined, like \eqr{eq:broad_fd_conv_main},
\begin{equation}
  \mathcal{F}_m^{\pm}(\varepsilon_0) = \frac{\Gamma}{\pi} \int_{-\infty}^{\infty} ~~~ \frac{f(\mp\epsilon)}{\Gamma^2 + ((\epsilon - \varepsilon_0)/\hbar - m \omega)^2} 
~~~d\epsilon  
\label{eq:broad_fd_conv}
\end{equation}
\noindent 
being the convolution of the Fermi-Dirac distribution and the effective Lorentzian density of states of the QD caused by the coupling with the CR. 
Here, we have disregarded the well-known diverging Lamb shift arising from taking the trace of the (unbound) CR number operator \cite{Sowa_Lambert_Seideman_Gauger_2020,hartung2022zetaregularized}.
Notably, in \eqr{eq:Gamma_pm_scq_long} appear the semiclassical phase of the driven level $\exp\left(-i\int_0^t dt' \varepsilon(t')/\hbar\right)$, as well as an expansion on the Floquet modes of the system (see Section.~\ref{Floquet_Modes}), recalling the functional form expected from Floquet-Lindblad theory \cite{Ikeda_Chinzei_Sato_2021,Mori_2023}.

Computationally, it is easier to explicitly write the tunnelling rate in terms of its Fourier components. Therefore, after another application of the Jacobi-Anger identity, \eqr{eq:Gamma_pm_scq_long} reads
\begin{equation}
  \begin{aligned}
  \Gamma_-(t) = \Gamma \sum_{m= - \infty}^{+ \infty} \sum_{n= - \infty}^{+ \infty} & J_n \left( \frac{\delta \varepsilon}{\hbar \omega}\right) J_m \left( \frac{\delta \varepsilon}{\hbar \omega}\right) \cdot\\ & \cdot \Re\left[ 
e^{ -i (m-n)\omega t}\mathcal{F}_m^{-}(\varepsilon_0)\right],
\end{aligned}
\label{eq:Gamma_scq_fourier}
\end{equation}
\noindent
which is the result presented in Section~\ref{self_ME}.

\section{Real and Virtual Photon Exchange}
\label{sect_GC}

In the main text, we have mostly discussed the absolute value of the SEB admittance, especially the maximum value and FWHM, for they are easily experimentally accessible properties. We believe of value, however, to explore more in depth the real and imaginary part of $Y$, introduced in Section~\ref{Floquet_Modes}, for they have very different \textit{electrical} (and physical) effects \cite{Vigneau_2023}. 
For simplicity, in this section we shall uniquely discuss the fundamental $Y_1$, dropping the suffix from now on, but this discussion is trivially extendable to the other harmonics \cite{oakes2022quantum}. Moreover, being in the small-signal regime, the effects of PLB are not considered in this section and the PhO is to be considered lossless.

To begin with, the presence of only first time derivatives in the LME in \eqr{eq:ME_gen} means that we can, without loss of generality, write
\begin{equation}
  Y = G_S + i \omega C_T,
\end{equation}
\noindent
where $C_T$ represents the \textit{tunnelling} capacitance while $G_S = 1 / R_S$ is the Sisyphus conductance \cite{Vigneau_2023,Mizuta2017}. 
We shall stress the notable absence of what is commonly referred to as \textit{quantum} capacitance \cite{Esterli2019}, for that corresponds to an adiabatic redistribution of charges along a unitary dynamics because of an avoided crossing \cite{Vigneau_2023,Mizuta2017}. The level crossing in a SEB at $\varepsilon=0$, however, is between two states describing the full or empty QD. Therefore, they formally belong to \textit{different} canonical single-body Hilbert spaces, and thus the crossing \textit{cannot} be avoided. 

Redistribution of charge, however, can still happen through \textit{incoherent} (stochastic) processes via the CR through the jump operators in the LME and may occur either elastically or inelastically.
In a cyclostationary process, this results in a gate current either in-phase or out-of-phase with the driving voltage. 
In the small-signal regime, this simply reads \cite{Esterli2019,oakes2022quantum,Mizuta2017}
\begin{equation}
  I_g = Y V_{in}.
\end{equation}
This semiclassical picture is of particular interest because it clearly shows how only the former, described by $G_S$, can lead to energy dissipation, while the signal arising from $C_T$ conserves energy over an rf cycle. 

We can picture these two types of response from a QED perspective as the QD mediating an interaction between the PhO and the CR. The capacitive response, thus, would correspond to the QD and the PhO exchanging \textit{virtual} photons, whose energy is not accounted for, in the form of creating a polaron and populating the Floquet ladder, while the resistive components will describe \textit{real} photons, which transfer energy to the PhO from the CR and vice versa.
Consisting of fully elastic jumps, the capacitive component is time-reversible. Thus, the $m$-th and $-m$-th Floquet modes must share the same occupation (i.e., \eqr{eq:Floquet_DOS}) and a precise phase relation. Stochastically, however, an \textit{extra} real photon can be emitted (absorbed) resulting in scattering between the Floquet modes. This imbalance in the $m$-th and $-m$-th modes manifests mathematically as a phase delay in the averaged steady-state response, and thus produces a resistive behaviour.

\begin{figure}[htb!]
  \centering
  \includegraphics[width = 0.99 \linewidth ]{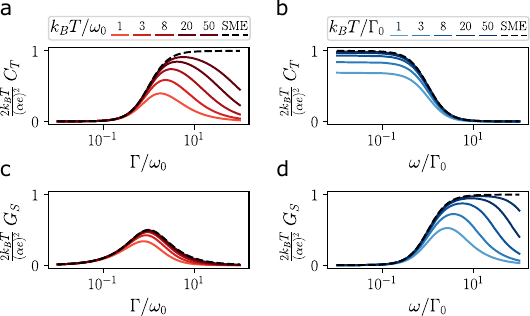}
  \caption{ Small-signal Tunnelling capacitance ($C_T$) and Sisyphus conductance ($G_S$) for varying $\Gamma$ and $\omega$. Panels (a,c) show how the effect of LB is mostly on the reactive component, while panels (b,d) shows how the main effect of FB is to lower the Sisyphus conductance. This is to be expected when considering energy dissipation in the CR.}
  \label{fig:fig5_GC}
\end{figure}

In light of this interpretation, it is interesting to consider the behaviour in Fig.~\ref{fig:fig5_GC}, which shows the (maximum of) tunnelling capacitance and Sisyphus conductance for increasing $\Gamma$ and $\omega$.

Semiclassically, increasing $\Gamma$ always leads to an increased signal (dashed lines in Figs.~\ref{fig:fig5_GC}a-c), as it increases the probability of tunnelling events per unit time. Moreover, for very large $\Gamma$ the signal is mostly capacitive, as, in the Instantaneous Eigenvalues Approximation, $G_S$ derives from electrons that \textit{failed to tunnel} elastically at the CR Fermi energy, but still have non-zero tunnelling probability because of thermal smearing.
The last point, as shown in Fig.~\ref{fig:fig5_GC}c, remains valid in the Floquet picture, as from the Floquet-Fermi golden rule we expect the CR to efficiently generate transitions between modes when the energy scale of its interaction ($h \Gamma$) is comparable with the photon energy \cite{Rudner_Lindner_2020}. For $\Gamma$ largely detuned from $\omega$ we expect the Floquet dynamics to be mostly unperturbed, and thus the admittance be mostly reactive. 
This has a simple interpretation in light of the time-energy Heisenberg uncertainty principle, recalling that tunnelling in and out of the QD requires the system to create and destroy photons. When the lifetime of the polaron becomes shorter than the photon energy, said photons can avoid being \textit{accounted for}, and thus this process becomes dissipationless. 
Because of LB, however, the Floquet dynamics consists of \textit{decaying} states, as obvious by considering the eigenstates of the self-consistent propagator in \eqr{eq:self_consistent_time_evol}.
When $\Gamma \gg \omega$, essentially \textit{all} dressed states will decay within an rf cycle. Therefore, we expect also the capacitive component to decay exponentially with the Floquet dynamics.

A similar point can be made about FB when we increase $\omega$ for fixed $\Gamma$ (Figs.~\ref{fig:fig5_GC}b,d). Semiclassically, we expect most of the signal to be resistive for large $\omega$, for the QD cannot keep up with the rf and thus the tunnelling phase will be largely randomized. However, we still expect the Sisyphus conductance to monotonically increase, as already argued in Section~\ref{Polaron_Broad}. This ought to be clear by the fact that a damped system which is driven too fast classically \textit{lags} 90\textdegree~out of phase with the driving force.
This, however, is not allowed by FB, as shown in Fig.~\ref{fig:fig5_GC}d. The quantum Zeno argument made in Section~\ref{Polaron_Broad} finds its natural counterpart in the description Floquet modes. If the coherent dynamics, in fact, becomes faster than the interaction with the CR, we expect the Floquet evolution to be mostly unperturbed. Floquet modes are, however, orthogonal over an rf cycle. Therefore, no transitions between dressed states becomes allowed, and the resistive signal drops. Perhaps more simply, this has to occur as for $\omega \gg \Gamma$ the evolution of the SEB becomes almost unitary, and thus it \textit{cannot} dissipate any energy.
We can make sense of this also from a polaron perspective, as increasing $\omega$ also increases the energy of the photons that are necessary to create the polaron. If $\Gamma < \omega$, moreover, those photons live longer than the Uncertainty Principle allows for virtual photons. Therefore, there must be energy transferred to and from the CR. If we are in the FB regime, however, this energy is larger than the average thermal energy $k_B T$ of the CR. This, therefore, will be far less likely to be able to provide \textit{real} photons to cause transitions between the modes. Thus the resistive signal will decrease.

Finally, we shall point out how, perhaps unsurprisingly, the SCQME deviates from the SME when $k_B T \approx \hbar \omega, h \Gamma$, while the semiclassical result is retained in the limit of large temperature.

\section{Optimal Small-Signal Parameters}
\label{Optimal_signal}

In this section, we will briefly discuss the combined effect of LB and FB on SEB readout. From Section~\ref{Polaron_Broad}, in fact, one could assume that increasing $\omega$ would be detrimental past a certain point because of FB, and similarly for $\Gamma$ because of LB.
While this is certainly true, however, this assumes that \textit{all} other parameters are fixed. 
If both $\Gamma$ and $\omega$ become free variables, which may occur in systems with tunable couplings and resonators~\cite{Scarlino_2022}.

\begin{figure}
  \centering
  \includegraphics[width = 0.99 \linewidth ]{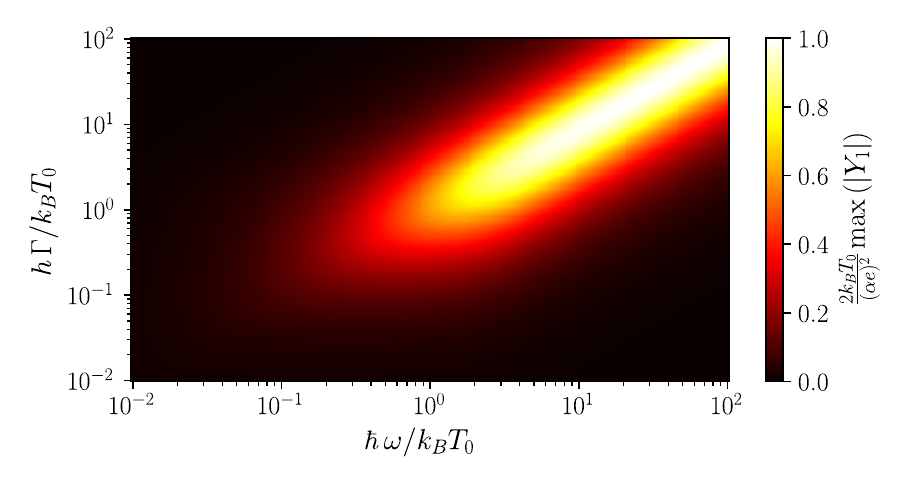}
  \caption{Maximum of the small-signal admittance $|Y_1|$ when varying $\Gamma$ and $\omega$. The figure shows how the maximum signal is always achieved when $\Gamma \approx \omega$, and both are as large as possible.}
  \label{fig:figMax}
\end{figure}

In Figure~\ref{fig:figMax} we show the impact of the small-signal admittance $|Y_1|$ at constant $k_B T$ as a function of $\Gamma$ and $\omega$. As it ought to be clear, the signal keeps increasing even if $k_B T \ll h \Gamma, \hbar \omega$, well beyond the semiclassical regime. 

Despite the dynamics now being fully dominated by quantum degrees of freedom of the QD dynamics, the semiclassical picture is still insightful. Increasing the frequency, in fact, means that we can get more tunnelling events per unit time. This occurs, however, only if the electrons can keep up with the fast oscillations. Therefore, that the signal is maximized when $\omega \sim \Gamma$, as transpires from Fig.~\ref{fig:figMax}.
This fact can also be explained in the quantum picture by the uncertainty principle. LB, in fact, arises from the finite lifetime of electrons in the QD. In practice, however, this is not \textit{seen} by the system if the QD is, on average, filled and emptied by the PhO \textit{faster} than the intrinsic lifetime of the level.
A similar argument can be made for FB, which can essentially be seen as a quantum Zeno phenomenon, as the electron gets trapped in the QD. A short enough lifetime, however, ensures that tunnelling can still occur. 
Therefore, these two effects complement each other and the signal keeps increasing at the onset of LB and FB, just more slowly than it would have done in the semiclassical model.

\section{From Admittance to Reflectometry Lineshape}
\label{lineshape}

\begin{figure}[h!]
  \includegraphics[width = \linewidth]{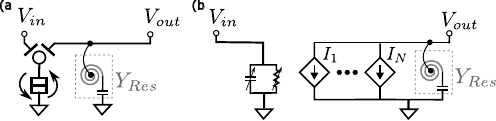}
  \caption{(a) Transmission experiment setup with an SEB connected in parallel with a resonator. (b) Equivalent circuit schematic in which the SEB is seen by the resonator as a set of VCCSs in parallel.}
  \label{fig:figRes}
\end{figure}

Having obtained the effective quantum impedance of the SEB, we need to compute the system's transmission coefficient $\mathcal{T} (Y_N)$ as a function of $Y_N$ when the system is coupled via the gate to the resonator, whose centre frequency is $\omega_{res} = N \omega$. 
It is easier to consider the case of the setup in Fig.~\ref{fig:figRes}a, with the SEB in parallel to the resonator. 

We can now define $\mathcal{T} (Y_N)$ by considering the phasor relation 
\begin{equation}
  V_{out}^N = \mathcal{T}(Y_N) V_{in}.
\end{equation}
\noindent
where $V_{out}^N(t) = \Re\left[V_{out} e^{i N \omega t}\right]$ and $V_{in}(t) = \Re\left[V_{in} e^{i \omega t}\right]$. We point out that, for $N=1$, this is nothing but the standard definition of the transmission coefficient.

To build an equivalent circuit model, we can now consider that the impedance seen by $V_{in}$ will be much larger than that of the line $Z_0 = 50 \Omega$, of the order of the quantum of resistance $e^2/h$. Therefore, most of the signal will be reflected.
For simplicity, we will assume that the drive is large enough to be able to neglect the self-loading of the collection gate \cite{cochrane2022intrinsic}.
In this case, the dynamics of the SEB as seen by the output can be replaced by a Voltage-Controlled Current Source (VCCS), whose output will be $I_N = \Re\left[V_{in} Y_N e^{i N \omega t}\right]$. 
Considering this equivalent circuit, we easily obtain

\begin{equation}
  V_{out}^N = - \frac{Y_N}{Y_{Res}} ~V_{in}.
\end{equation}

We note that the impedance of the resonator, in parallel to the VCCS, will be significant only at the resonant frequency, while it shall act as a short to ground at all other frequencies. This justifies the formalism of equivalent admittance, since all the terms not oscillating at the resonant frequency (here taken $\omega_{Res} = N \omega$) are short to ground. 
For the resonant term, however, we see how the low admittance of the resonator amplifies the resonant signal. Moreover, we can see how the effective admittance of the SEB, or rather, its \textit{transconductance}, is directly proportional to the transmitted signal.
This last observation allows us to directly relate Eqs.\ (\ref{eq:Y_sc}) and (\ref{eq:Y_q}) to experimental reflectometry data.

The high admittance of the off-resonance resonator forces us to use 2 separate gates for the driving. For simplicity, we imagine them to have negligible crosstalk and the same lever arm. If this is not the case, one can simply account for this by modifying $\alpha^2 \rightarrow \alpha_d \alpha_c$ in $\mathcal{C}_N$ (\eqr{eq:CN}), where $\alpha_d$ and $\alpha_c$ are the lever arms of the driving and collection gates respectively \cite{oakes2022quantum}.

Lastly, we point out that, if the measurement is made with a \textit{single} gate in reflection rather than transmission, by a similar argument one can immediately write
\begin{equation}
  V_{out}^{\mathcal{R}} = \left(1 - \frac{Y_1}{Y_{Res}}\right)~V_{in}
\end{equation}
\noindent
where we highlight that only the fundamental can be observed through the resonator in this case.
 
\section{The broadened Fermi-Dirac}
\label{Digamma}

In this section, we will briefly discuss the mathematical steps that lead to the digamma function in \eqr{eq:digamma}. While this result allows for some physical insights (e.g., the discussion of \eqr{eq:quasienergies}), a closed-form equation for $\mathcal{F}_m$ allows for dramatically lower computation times. In particular, the expansion of the digamma (and trigamma) functions as infinite series or continued fractions allows for zero-cost inclusion of LB in QD simulations.

To begin with, we notice the obvious fact that
\begin{equation}
  \mathcal{F}_m^{\pm}(\varepsilon_0) = \frac{\Gamma}{\pi} \int_{-\infty}^{\infty} ~~~ \frac{f(\pm\epsilon)}{\Gamma^2 + ((\epsilon - \varepsilon_0)/\hbar - m \omega)^2} 
~~~d\epsilon  
\end{equation}
\noindent
is a convolution between the Fermi-Dirac and a Lorentzian, and thus can be computed as a product in reciprocal space. In the following, we shall for simplicity only consider the case of $\mathcal{F}_0^{-}(\varepsilon_0)= \mathcal{F}(\varepsilon_0)$, and all other cases trivially follow.

Before tackling the problem, we can simplify the calculation by noticing that 
\begin{equation}
  f(\epsilon) = \frac{1}{2} - \frac{1}{2} \tanh{\left(\frac{\epsilon}{2 k_B T}\right)}
  \label{eq:Fermi_app}
\end{equation}
\noindent
and thus
\begin{equation}
  \mathcal{F}(\varepsilon_0) = \frac{1}{2} - \frac{\Gamma}{2\pi} \int_{-\infty}^{\infty} ~~~ \frac{\tanh{\left(\frac{\epsilon}{2 k_B T}\right)}}{\Gamma^2 + (\epsilon - \varepsilon_0)^2/\hbar^2} ~~~d \epsilon.
\end{equation}
We can now make use of the fact that the Fourier transform of a Lorentzian is a decaying exponential, while,
\begin{equation}
  \mathfrak{F}\left[\tanh{\left(\frac{\epsilon}{2 k_B T}\right)}\right](\xi) = \frac{i \pi k_B T}{\sinh{\pi k_B T \xi}}
\end{equation}
\noindent 
where $\mathfrak{F}$ indicates the Fourier transform in a distribution sense, to write
\begin{equation}
  \mathcal{F}(\varepsilon_0) = i ~ \mathfrak{F}^{-1} \left[\frac{\pi k_B Te^{- |\Gamma \xi|}}{\sinh{(\pi k_B T \xi)}}\right].
\end{equation}

We can now use the fact that $\mathcal{F}(\varepsilon_0) - \frac{1}{2}$ is antisymmetric to write the integrals as 
\begin{equation}
  \mathcal{F}(\varepsilon_0) - \frac{1}{2} = i \pi k_B T \left(\int_0^{\infty} \frac{e^{i \varepsilon_0 \xi - \Gamma \xi}e^{\pi k_B T \xi}}{1-e^{-2\pi k_B T \xi}} + c.c.\right).
  \label{eq:almost_digamma}
\end{equation}
To get the desired result we must now perform the substitution $t = 2\pi k_B T \xi$ and notice that, for $\Re[z]>0$, we can write the digamma function as\cite{whittaker_watson_1996}
\begin{equation}
  \psi_0(z) = \int_0^{+\infty} \left(\frac{e^{-t}}{t} - \frac{e^{-zt}}{1-e^{-t}}\right) dt.
\end{equation}

Adding and subtracting $\frac{e^{-t}}{t}$ to \eqr{eq:almost_digamma} therefore, we get 
\begin{equation}
  \begin{aligned}
  \mathcal{F}(\varepsilon_0) - \frac{1}{2} = \frac{1}{2\pi}\bigg(\psi_0\left(\frac{1}{2} + \frac{h\Gamma  + i \varepsilon_0}{2 \pi k_B T}\right) - 
  \\ - \psi_0\left(\frac{1}{2} + \frac{h\Gamma - i\varepsilon_0}{2 \pi k_B T}\right)\bigg).
\end{aligned}
\end{equation}

Finally, we can use the identity
\begin{equation}
  \psi_0(z^*) = \psi_0(z)^*
\end{equation}
\noindent
to write 
\begin{equation}
  \mathcal{F}(\varepsilon_0) = \frac{1}{2} - \frac{1}{\pi} \Im\left[\psi_0\left(\frac{1}{2} + \frac{ h\Gamma  + i \varepsilon_0}{2 \pi k_B T}\right)\right]
  \label{eq:digamma_app}
\end{equation}
\noindent
of which \eqr{eq:digamma} is an immediate generalization.

As a final remark, an interesting sanity check is that it is a well-known fact from taking the logarithmic derivative of the Gamma function that, for $x \in \mathbb{R}$,
\begin{equation}
  \Im\left[\psi_0\left(\frac{1}{2} + x\right)\right] = \frac{\pi}{2} \tanh{(\pi x)}
\end{equation}
\noindent
which immediately shows how \eqr{eq:digamma_app} retrieves \eqr{eq:Fermi_app} in the limit of $\Gamma =0$.

\section{Polaron-Transformed Hamiltonian}
\label{H_pol}

In this Appendix, we discuss the mathematical manipulations presented in Section~\ref{self_ME}. To make it more accessible to the reader, we will recall well-known mathematical properties when used.
We begin by noticing that $[d,a] = [c,a] =0$, and similarly for any of the respective adjoints. Thus, we can simply picture the polaron transformation as a displacement of the photonic state, which happens to be dependent on the QD occupation.

In the main text, we made reference to the BCH theorem, which states for two matrices $A$ and $B$ that 
\begin{equation}
  e^A B a^{-A} = B + [B, A]
  \label{eq:BCH_app}
\end{equation}
\noindent 
if $[B, A]$ is a \textit{c}-number (i.e., $[B, A]$ commutes with both $A$ and $B$).
Considering this property, for any displacement operator $D_\alpha = \exp(\alpha a^\dagger  - a^* a)$, we have
\begin{equation}
  D_\alpha a D\alpha^\dagger = a + \alpha,
    \label{eq:displacement_shift}
\end{equation}
\noindent
from which it follows immediately that \cite{Sowa_Lambert_Seideman_Gauger_2020,Sowa_Lambert_Seideman_Gauger_2020}
\begin{equation}
  e^{-S} \parens{H_{PhO}+ H_{DP}} e^{S} =
  \frac{g^2}{\hbar \omega} d^\dagger d + \hbar \omega a^\dagger a.
\end{equation}

This is particularly remarkable because, referencing Section~\ref{Semiclassical_gamma}, the term $a + a^\dagger$, which gives the semiclassically oscillating QD energy has now disappeared. Moreover, that term obviously contains rotating and counter-rotating waves. Thus, especially recalling the discussion in Section~\ref{Power_Broad}, it ought to be clear how a rotating-wave approximation in the usual Floquet-Rabi sense is generally not possible if we want to describe effective admittances. 

Equation~(\ref{eq:HDR_pol}) could be derived directly from \eqr{eq:BCH_app} using fermionic commutation relations. However, a simple trick is to note that $a - a^\dagger$ is a Grassmann number. Therefore, $e^S$ is \textit{also} a fermionic displacement operator, for which \eqr{eq:displacement_shift} is also valid\cite{Cahill_Glauber_1999}.

Therefore, we have shown that the polaron transformation is \textit{simultaneously} displacing both the PhO depending on the state of the QD \textit{and} the QD because of the PhO electric field. This characteristic of double displacement allows us to completely remove the semiclassically oscillating field but also invites us to think of the problem no more as electrons interacting with photons but as a \textit{combined} quasiparticle (hence the name polaron). 
This picture is retrieved if we consider the QD-CR interaction in the polaron frame (\eqr{eq:HDR_pol}). An interaction of the form 
\begin{equation}
  c_\epsilon D^\dagger d^\dagger
\end{equation}
\noindent
destroys a fermion in the CR and simultaneously creates an electron in the QD \textit{and} displaces the PhO to create a polaron.

Finally, we shall note that, technically, the rates in \eqr{eq:Gamma_scq_fourier} are obtained in a different frame than the semiclassical description. However, from \eqr{eq:LME_mat}, we see how, at the steady state, $\rho$ is purely diagonal. Therefore, the density operator reads
\begin{equation}
  \rho_{SS}(t) = \frac{1}{2} \mathbb{I} + \left(P(t) - \frac{1}{2}\right) \sigma_z
\end{equation}
\noindent
where $\mathbb{I}$ is the identity and, obviously, $\sigma_z = d^\dagger d$ in the two-level picture. Thus, $[\rho_{SS}(t), S] = 0$, and the occupation probability in the polaron frame is the same as in the lab frame. This is an obvious consequence of $\ket{e}$ and $\ket{o}$ belonging to \textit{different} canonical Hilbert spaces, and thus they cannot be mixed by the canonical Lang-Firsov transformation.

\bibliographystyle{unsrtnat}

\end{document}